\documentclass[aps,prd,twocolumn,nofootinbib,longbibliography,10pt]{revtex4-2}
\usepackage{etoolbox}
\usepackage{dcolumn}
\usepackage{amsmath,amssymb,amsfonts,mathtools}
\usepackage{mathrsfs}
\usepackage{graphicx,tensor}
\RequirePackage[colorlinks=true,urlcolor=blue,citecolor=red,linkcolor=blue]{hyperref}
\usepackage{natbib,bbold}
\usepackage{BOONDOX-cal}
\usepackage{wrapfig}
\usepackage{flushend}
\usepackage{nicefrac,xfrac}

\makeatletter
\newsavebox{\@brx}
\newcommand{\llangle}[1][]{\savebox{\@brx}{\(\m@th{#1\langle}\)}%
  \mathopen{\copy\@brx\kern-0.5\wd\@brx\usebox{\@brx}}}
\newcommand{\rrangle}[1][]{\savebox{\@brx}{\(\m@th{#1\rangle}\)}%
  \mathclose{\copy\@brx\kern-0.5\wd\@brx\usebox{\@brx}}}
\makeatother
\begin{document}
\title{Quantum nature of gravity in a Bose-Einstein condensate}
\author{Soham Sen}
\email{sensohomhary@gmail.com, omthakurma.sensohom@gmail.com}
\affiliation{Department of Astrophysics and High Energy Physics, S. N. Bose National Centre for Basic Sciences, JD Block, Sector-III, Salt Lake City, Kolkata-700 106, India}
\author{Sunandan Gangopadhyay}
\email{sunandan.gangopadhyay@gmail.com}
\affiliation{Department of Astrophysics and High Energy Physics, S. N. Bose National Centre for Basic Sciences, JD Block, Sector-III, Salt Lake City, Kolkata-700 106, India}
\begin{abstract}
\noindent The effect of noise induced by gravitons on a Bose-Einstein condensate has been explored in \href{https://link.aps.org/doi/10.1103/PhysRevD.110.026014}{Phys. Rev. D 110 (2024) 026014}. In the previous paper, we investigated the effects of graviton while detecting a gravitational wave using a Bose-Einstein condensate. In this work, we shall explicitly calculate the decoherence due to the noise of gravitons for a maximally entangled momentum states of the Bose-Einstein condensate. This decoherence happens due to Bremsstrahlung from the Bose-Einstein condensates due to the effect of the noise induced by gravitons. It is observed that the maximally entangled state becomes entangled with the graviton state and it decays over time as a result of this gravitational Bremsstrahlung. This new entangled state is termed as a Bose-Einstein supercondensate. Using this property of the Bose-Einstein condensate in a quantum gravity background, we propose an experimental test via the use of atom lasers (generated from the condensate) which would, in principle, help to detect gravitons in future generations of very advanced ultra-cold temperature experiments.
\end{abstract}
\maketitle
\section{Introduction}
\noindent The quantum nature of gravity is one of the most precious pearls hidden deep inside the depths of the cosmic ocean. It is already evident that even at the quantum mechanical regime gravity behaves classically.  The fundamental reason for searching for a quantum theory of gravity is that at regimes like the Big Bang singularity or the singularity of a black hole, the general theory of relativity breaks down. It is assumed that at such small length scales a different theory, namely ``quantum gravity" will be dominant. Although a claim was made in \cite{Singularity_Free_Theories_Gravity} which states that at such small regimes, gravity becomes sufficiently weak so that other fundamental forces are dominant. Another aspect is that at larger length scales quantum theory breaks down and gravity is, by all its entirety, a purely classical field \cite{Karolyhazy,Penrose,Diosi,Bassi,Bassi2}. However, the altruistic drive to detect the quantum nature of gravity lies in the fact that all three other fundamental forces of nature can be described by a quantum field theory. A fascinating effect regarding the detection of the quantum nature of gravity was proposed initially by Bronstein \cite{Gorelik} and later by Feynman \cite{Feynman} in the Chapel Hill conference \cite{Chapel_Hill}. The initial theoretical framework for a linearized quantum theory of gravity was developed by Bronstein \cite{Bronstein1,Bronstein2} and later in \cite{SNGupta1,SNGupta2} it was analyzed that the quanta of linearized gravity theory, a ``graviton", is either a spin two or a spin-zero particle. The evolution of the quantum gravity formalism in the first half of the twentieth century was encaptured in details in \cite{BlumRickels}. Linearized quantum gravity theory being a low-energy theory is very hard to detect in nature as the signatures from matter-field interactions are very weak and nearly impossible to detect using current experimental techniques. There recently has been a claim regarding unearthing the graviton signatures using the entanglement generated between two massive bodies via graviton interaction \cite{Bose,Marletto,Marletto2,Bose2,Bose3}. Because two gravitons can superpose (as linearized quantum gravity theory obeys the quantum superposition principle), a coherent matter source can act as a viable candidate for proving the quantum nature of gravity \cite{Marletto,Marletto2,Bose,Bose2}. The primary challenge of such an experimental set up is creating a coherent mass-dual. We have also come to know about a very recent work that involves the use of matter-wave interferometry to detect dipole-dipole decoherence rate of the quantum-gravity induced entanglement in masses\cite{Bose_Interferometry}. Recently, in \cite{Haine}, the possibility of detecting quantum gravity signatures in quantum gases has been explored. Very recently, there have been several investigations regarding the conceptualization of another aspect of linearized quantum gravity theory. Making use of a simple two-point gravitational wave detector model one can show using a path integral approach that the geodesic deviation equation becomes Langevin-like because of stochastic inputs coming from the noise induced by gravitons \cite{QGravNoise,QGravLett,QGravD,AppleParikh,OTMGraviton,OTMApple}. In \cite{OTMGraviton}, it has been shown that such interactions tamper with the Heisenberg uncertainty relation leading it to a structure that imitates the generalized uncertainty principle in the Planck-mass limit. Using a canonical approach the same effect has been observed in a 3+1-dimensional model in \cite{KannoSodaTokuda} and later this point particle-graviton interaction model was used to entangle two perpendicular mirrors of a gravitational wave detector (at separate arms) and decoherence due to gravitational bremsstrahlung was obtained \cite{KannoSodaTokuda2} which may be measurable in a very advanced future experimental scenario. 

\noindent Another very important aspect of low-temperature theoretical as well as experimental physics is a Bose-Einstein condensate  \cite{SNBose,Einstein1,Einstein2,Einstein3,1Nobel2001,2Nobel2001}. There have been several proposals regarding the use of a Bose-Einstein condensate in gravitational wave detection \cite{PhononBEC,PhononBEC2,PhononBEC3,PhononBEC4,
ThesisMatthew}. Recently in \cite{Super_Condensate_OTM,Super_Condensate_OTM_Lett}, we explored the case of a gravitational wave interacting with a Bose-Einstein condensate when the gravitational fluctuation is quantized to include graviton interaction in the theory. We have made use of the quantum Fisher information techniques to see the signatures of quantization of gravity when the Bose-Einstein condensate is in resonance with the incoming gravitational wave. We observe that the pseudo-Goldstone bosons get infused by the noise of gravitons and the corresponding equation of motion becomes a Langevin-like stochastic second-order differential equation. The primary detection scenario comes from the fact that even at initial times there is a finite detection probability for the gravitons because of the existing fluctuation field all around the Bose-Einstein condensate (BEC). In this paper, we have extended this work to a much more detailed detection scenario using atom lasers and atom interferometry. In this second work, we make use of the equation of motion for the time-dependent part of the pseudo-Goldstone boson and its corresponding solution from \cite{Super_Condensate_OTM,Super_Condensate_OTM_Lett} and use it in this analysis. Using the Liouville equation for the total density matrix and tracing out the field degrees of freedom, it is observed that between two separate modes, which are the eigenstates of the transverse wave-number operators,  there is decoherence due to gravitational Bremsstrahlung. We propose a new kind of experimental set up using atom lasers from a Bose-Einstein condensate. We also observe that due to the interaction of the gravitons with a BEC, the single-mode phonon states become entangled which gives rise to a new kind of \textit{supercondensate} structure of the Bose-Einstein condensate. This entanglement generation explains why there was a non-vanishing initial-time detection probability present while making use of the quantum gravitational Fisher information in \cite{Super_Condensate_OTM,Super_Condensate_OTM_Lett}\footnote{Some important aspects of Bose-Einstein condensates in cosmological scenario has been explored in \cite{DymnikovaKhlopov,DymnikovaKhlopov2}. Recently in \cite{SpaceBased}, space-based optical lattice clocks has been proposed as a new gravitational wave detector.}.

\noindent  The paper is organized as follows. In section (\ref{S2}), we recall the results from our first work (\cite{Super_Condensate_OTM,Super_Condensate_OTM_Lett}) and use the solution of the time-dependent part of the pseudo-Goldstone bosons. In section (\ref{S3}), we calculate the gravitational Bremsstrahlung using the Liouville equation via tracing out the field degrees of freedom. Finally, in section (\ref{S4}), we propose an experimental set up for detecting the signature of linearized quantum gravity due to graviton-induced Bremsstrahlung from a Bose-Einstein ``supercondensate". Finally in section (\ref{S5}), we summarize our results.
\section{Important results from the ``first treatise"} \label{S2}
\noindent In the first part of our analysis \cite{Super_Condensate_OTM,Super_Condensate_OTM_Lett}, we have considered a self-interacting complex scalar field theory with $\lambda|\phi|^4$ interaction term, interacting with a gravitational fluctuation over a flat background with the fluctuations being quantized to imitate gravitons\footnote{We have made use of the natural units, $\hbar=c=1$, throughout our analysis and later have restored them.}. The background can be expressed as 
\begin{equation}\label{2.1}
g_{\mu\nu}=\eta_{\mu\nu}+h_{\mu\nu}
\end{equation}
where $\eta_{\mu\nu}=\text{diag}\{-1,1,1,1\}$. The gravitational fluctuation term in the transverse traceless gauge can be recast as $h_{\mu\nu}=\bar{h}_{\mu\nu}+\partial_\mu\xi_\nu+\partial_\nu\xi_\mu$ and one can make use of the discrete-mode decomposition given by
\begin{equation}\label{2.2}
\bar{h}_{ij}(t,\mathbf{x})=\frac{2\kappa}{\sqrt{V}}\sum_{\mathbf{k},s}h^s(t,\mathbf{k})e^{i\mathbf{k}\cdot\mathbf{x}}\epsilon^s_{ij}(\mathbf{k})
\end{equation}
with $\kappa=\sqrt{8\pi G}$. The complex scalar field is expressed as
\begin{equation}\label{2.3}
\phi(t,\mathbf{x})=e^{i(-\tilde{\sigma}t+\pi(t,\mathbf{x}))}\varphi(t,\mathbf{x})
\end{equation}
where $\varphi(t,\mathbf{x})$ denotes the heavy field and $\pi(t,\mathbf{x})\in\mathbb{R}$ denotes the pseudo-Goldstone bosons. Integrating out the heavy fields from the theory and combining it with the Einstein-Hilbert action in the transverse-traceless gauge, we obtain the total action for the model system as
\begin{equation}\label{2.3a}
\begin{split}
S=&-\frac{1}{8\kappa^2}\int d^4 x~ \partial_{\kappa}\bar{h}_{ij}\partial^{\kappa}\bar{h}^{ij}+\gamma_\beta\int dt\biggr[\sum\limits_{\mathbf{k}_\beta}\bigr|\dot{\psi}_{\mathbf{k}_\beta}(t)\bigr|^2\\&-c_s^2\Bigr[\eta_{ij}+\bar{h}_{ij}(t,0)\Bigr]\sum\limits_{\mathbf{k}_\beta}k_\beta^ik_\beta^j\bigr|\psi_{\mathbf{k}_\beta}(t)\bigr|^2\biggr]
\end{split}
\end{equation}
where $\gamma_\beta=\frac{V_\beta}{2\lambda}(3\tilde{\sigma}^2-m^2)$ and $c_s^2=\frac{\tilde{\sigma}^2-m^2}{3\tilde{\sigma}^2-m^2}$\footnote{If a proper dimensional reconstruction is done then $\kappa=\sqrt{\frac{8\pi\hbar G}{c^3}}$, $\gamma_\beta=\frac{V_\beta}{2\lambda c}\left(\frac{3\tilde{\sigma}^2}{c^2}-\frac{m^2c^2}{\hbar^2}\right)$, and $c_s^2=c^2\frac{\left(\nicefrac{\tilde{\sigma}^2}{c^2}-\nicefrac{m^2c^2}{\hbar^2}\right)}{\left(\nicefrac{3\tilde{\sigma}^2}{c^2}-\nicefrac{m^2c^2}{\hbar^2}\right)}$.} which denotes the square of the speed of sound. Here, $V_\beta$ denotes the volume of the box in which the phonons are being quantized. Making use of the mode decomposition given in eq.(\ref{2.2}) and substituting it in the above action, we can recast the total action for the system as
\begin{equation}\label{2.4}
\begin{split}
S=&\frac{1}{2}\sum_{\textbf{k},s}\int dt\left(\bigr|\dot{h}^s(t,\mathbf{k})\bigr|^2-k^2\bigr|h^s(t,\mathbf{k})\bigr|^2\right)+\\&\gamma_\beta\int dt\biggr[\sum\limits_{\mathbf{k}_\beta}\bigr|\dot{\psi}_{\mathbf{k}_\beta}(t)\bigr|^2-c_s^2\Bigr[\eta_{ij}+\frac{2\kappa}{\sqrt{V}}\sum\limits_{\mathbf{k},s}h_{\mathbf{k},s}(t)\\&\times\epsilon^s_{ij}(\mathbf{k})\Bigr]\sum\limits_{\mathbf{k}_\beta}k_\beta^ik_\beta^j\bigr|\psi_{\mathbf{k}_\beta}(t)\bigr|^2\biggr]~.
\end{split}
\end{equation}
Extremizing the action in eq.(\ref{2.4}) with respect to $\psi^*_{\mathbf{k}_\beta}(t)$, we obtain the following equation of motion \cite{Super_Condensate_OTM}
\begin{equation}\label{2.5}
\ddot{\psi}_{\mathbf{k}_\beta}(t)+c_s^2\Bigr[\eta_{ij}+\frac{2\kappa}{\sqrt{V}}\sum\limits_{\mathbf{k},s}h_{\mathbf{k},s}(t)\varepsilon^s_{ij}(\mathbf{k})\Bigr]k_\beta^ik_\beta^j\psi_{\mathbf{k}_\beta}(t)=0
\end{equation}
with $\varepsilon^s_{ij}(\mathbf{k})$ denoting the polarization tensor. Extremizing the action with respect to $h^*_{\mathbf{k},s}$, we obtain the equation of motion corresponding to the graviton mode as \cite{Super_Condensate_OTM}
\begin{equation}\label{2.6}
\begin{split}
\ddot{h}_{\mathbf{k},s}(t)+k^2h_{\mathbf{k},s}(t)=-\frac{4\gamma_\beta\kappa c_s^2}{\sqrt{V}}\epsilon^{s*}_{ij}(\mathbf{k})\sum\limits_{\mathbf{k}_\beta} k_\beta^ik_\beta^j\left\lvert\psi_{\mathbf{k}_\beta}(t)\right\rvert^2.
\end{split}
\end{equation}
In eq.(\ref{2.4}), the pseudo-Goldstone is decomposed into two parts $\pi(t,\mathbf{x})=\sum_{\mathbf{k}_\beta} e^{i\mathbf{k}_\beta \cdot\mathbf{x}}\psi_{\mathbf{k}_\beta}(t)$. It is straightforward to infer from the above decomposition that the $\psi_{\mathbf{k}_\beta}(t)$ term dictates the dynamical nature of the BEC and the spatial part has no dynamical contribution. The action in eq.(\ref{2.4}), can be expressed in terms of the Lagrangian of the full system as $S=\int dt L$ where the Lagrangian for the system is given by
\begin{widetext}
\begin{equation}\label{2.Lagrangian}
L=\frac{1}{2}\sum_{\textbf{k},s}\left(\bigr|\dot{h}^s(t,\mathbf{k})\bigr|^2-k^2\bigr|h^s(t,\mathbf{k})\bigr|^2\right)+\gamma_\beta\biggr[\sum\limits_{\mathbf{k}_\beta}\bigr|\dot{\psi}_{\mathbf{k}_\beta}(t)\bigr|^2-c_s^2\Bigr[\eta_{ij}+\frac{2\kappa}{\sqrt{V}}\sum\limits_{\mathbf{k},s}h_{\mathbf{k},s}(t)\epsilon^s_{ij}(\mathbf{k})\Bigr]\sum\limits_{\mathbf{k}_\beta}k_\beta^ik_\beta^j\bigr|\psi_{\mathbf{k}_\beta}(t)\bigr|^2\biggr]~.
\end{equation}
\end{widetext}
 From the form of the above Lagrangian, it is straightforward to write down the conjugate variables corresponding to $h_{\mathbf{k},s}(t)$ and $h^*_{\mathbf{k},s}(t)$ as
\begin{equation}\label{2.7}
\begin{split}
\mathfrak{P}_{\mathbf{k},s}(t)=\frac{\partial L}{\partial \dot{h}_{\mathbf{k},s}(t)}=\frac{1}{2}\dot{h}_{\mathbf{k},s}^{*}(t),~\mathfrak{P}^*_{\mathbf{k},s}(t)=\frac{1}{2}\dot{h}_{\mathbf{k},s}(t).
\end{split}
\end{equation}
Similarly for $\psi_{\mathbf{k}_\beta}(t)$ and $\psi^*_{\mathbf{k}_\beta}(t)$, the conjugate variables read
\begin{equation}\label{2.8}
P_{\mathbf{k}_\beta}(t)=\gamma_{\beta}\dot{\psi}^*_{\mathbf{k}_\beta}(t),~P^*_{\mathbf{k}_\beta}(t)=\gamma_{\beta}\dot{\psi}_{\mathbf{k}_\beta}(t).
\end{equation}
Using the above two equations, one can express the total Hamiltonian of the system as
\begin{equation}\label{2.9}
H(t)=H_0(t)+H_{\text{int}}(t)
\end{equation}
where the base Hamiltonian is given by
\begin{equation}\label{2.10}
\begin{split}
H_0(t)&=2\sum\limits_{\mathbf{k},s}\mathfrak{P}_{\mathbf{k},s}(t)\mathfrak{P}^*_{\mathbf{k},s}(t)+\frac{1}{2}\sum\limits_{\mathbf{k},s}k^2h_{\mathbf{k},s}(t)h^*_{\mathbf{k},s}(t)\\
&+\frac{1}{\gamma_\beta}\sum\limits_{\mathbf{k}_\beta}P_{\mathbf{k}_\beta}(t)P^*_{\mathbf{k}_\beta}(t)+\gamma_\beta c_s^2 \sum\limits_{\mathbf{k}_\beta}k_\beta^2\psi_{\mathbf{k}_\beta}(t)\psi^*_{\mathbf{k}_\beta}(t)
\end{split}
\end{equation}
and the interaction part of the Hamiltonian reads
\begin{equation}\label{2.11}
H_{\text{int}}(t)=\frac{2\gamma_\beta\kappa c_s^2}{\sqrt{V}}\sum\limits_{\mathbf{k},s}\sum\limits_{\mathbf{k}_\beta}h_{\mathbf{k},s}(t)\varepsilon^s_{ij}(\mathbf{k})k_\beta^ik_\beta^j\psi_{\mathbf{k}_\beta}(t)\psi^*_{\mathbf{k}_\beta}(t)~.
\end{equation}
We are considering a single mode for the BEC. Hence, for a single-mode BEC, one can get rid of the sum over the phonon modes in eq.(s)(\ref{2.10},\ref{2.11}). It is important to note that our primary aim is to explicitly observe the effect of gravitons on the phonon eigenstates corresponding to the transverse wave-number operator. To achieve this, we now consider the operator-raised form of the interaction Hamiltonian. This reads
\begin{equation}\label{2.12}
\hat{H}_{\text{int}}(t)=\frac{2\gamma_\beta\kappa c_s^2}{\sqrt{V}}\sum\limits_{\mathbf{k},s}\hat{h}_{\mathbf{k},s}(t)\varepsilon^s_{ij}(\mathbf{k})\sum\limits_{\mathbf{k}_\beta}\hat{k}_\beta^i\hat{k}_\beta^j\hat{\psi}_{\mathbf{k}_\beta}(t)\hat{\psi}^*_{\mathbf{k}_\beta}(t)
\end{equation}
where the form of the the graviton operator $\hat{h}_{\mathbf{k},s}$ is given by 
\begin{equation}\label{2.13}
\begin{split}
\hat{h}_{\mathbf{k},s}(t)=&h_{\text{cl}}^s(\mathbf{k},t)+\delta \hat{h}^I_{\mathbf{k},s}(t)-\frac{4\gamma_\beta\kappa c_s^2}{\sqrt{V}}\epsilon^{s*}_{ij}(\mathbf{k})\sum\limits_{{\hat{\mathbf{k}}}_\beta} {{\hat{k}}_\beta}^i{{\hat{k}}_\beta}^j\\&\times\int_0^t dt'\frac{\sin(k(t-t'))}{k}\bigr|\hat{\psi}_{\mathbf{k}_\beta}(t')\bigr|^2
\end{split}
\end{equation}
with the definition $\langle \hat{h}^I_{\mathbf{k},s}\rangle=h^s_{\text{cl}}(\mathbf{k},t)$ where the expectation of the graviton operator (in the interaction picture) is taken with respect to the initial state of the graviton.
Here, we are mostly focussed on the momentum states of the phonons as the Bose-Einstein condensate is formed in the momentum space via the superposition of the matter waves corresponding to the individual phonon modes.
\section{Decoherence due to graviton-BEC interaction by the emission of bremsstrahlung}\label{S3}
\noindent In this section, we shall calculate the decoherence happening due to the interaction of the BEC system with the gravitons. We shall follow here the method used in \cite{bremsstrahlung} to compute the decoherence between two momentum states of the BEC. Later this approach was adopted in \cite{KannoSodaTokuda,KannoSodaTokuda2}. We take the density operator of the BEC-graviton interaction system to be $\hat{\rho}$ and the corresponding Liouville super-operator is defined as
\begin{equation}\label{2.14}
\begin{split}
\hat{\mathfrak{L}}_{su}\hat{\rho}(t)&=-\frac{i}{\hbar}[\hat{H}(t),\hat{\rho}(t)]\\&=-\frac{i}{\hbar}[\hat{H}_0(t),\hat{\rho}(t)]-\frac{i}{\hbar}[\hat{H}_{\text{int}}(t),\hat{\rho}(t)]\\
\implies \hat{\mathfrak{L}}_{su}\hat{\rho}(t)&=\hat{\mathfrak{L}}_{0}\hat{\rho}(t)+\hat{\mathfrak{L}}_{\text{int}}\hat{\rho}(t)~.
\end{split}
\end{equation}
with $\hat{H}(t)$ being the total Hamiltonian of the system from eq.(\ref{2.9}) when only a single mode of the BEC is being considered. In the above equation, $\hat{\mathfrak{L}}_0$ denotes the Liouville superoperator corresponding to the base Hamiltonian $\hat{H}_0$ and $\hat{\mathfrak{L}}_{\text{int}}$ denotes the Liouville superoperator corresponding to the interaction part of the Hamiltonian in eq.(\ref{2.12}). Now the Liouville equation can be written as\begin{equation}\label{2.15}
\frac{\partial \hat{\rho}(t)}{\partial t}=-\frac{i}{\hbar}[\hat{H}(t),\hat{\rho}(t)]~.
\end{equation}
If at the initial time $t_i$, the form of the density matrix of the system is given by $\hat{\rho}(t_i)$ then using a recursion relation and making use of eq.(\ref{2.14}), one can obtain a solution of the density matrix of the entire system to be \cite{bremsstrahlung}
\begin{equation}\label{2.16}
\hat{\rho}(t)=\mathcal{T}\left[e^{\int_{t_i}^tdt'\hat{\mathfrak{L}}_{su}(t')}\right]\hat{\rho}(t_i)
\end{equation}
where $\mathcal{T}$ defines the time ordering. If we now trace over the field degrees of freedom corresponding to the graviton part, one is left with the density matrix corresponding to the BEC at some final time $t_f$ as
\begin{equation}\label{2.17}
\begin{split}
\hat{\rho}_{\text{BEC}}(t_f)&=\text{tr}_\mathcal{F}\left[\hat{\rho}(t_f)\right]\\
&=\text{tr}_\mathcal{F}\left[\mathcal{T}\Bigr[e^{\int_{t_i}^{t_f}dt'\hat{\mathfrak{L}}_{su}(t')}\Bigr]\hat{\rho}(t_i)\right]~.
\end{split}
\end{equation}
The time ordering can identically be separated into two parts, one corresponding to the BEC and the other corresponding to the graviton field as $\mathcal{T}=\mathcal{T}^{\mathcal{F}}\mathcal{T}^{\text{BEC}}$. Eq.(\ref{2.17})  can be recast using eq.(\ref{2.14}) as
\begin{widetext}
\begin{equation}\label{2.18}
\begin{split}
\hat{\rho}_{\text{BEC}}(t_f)&=\mathcal{T}^{\text{BEC}}\biggr[e^{\int_{t_i}^{t_f}dt'\hat{\mathfrak{L}}_0(t')}\text{tr}_{\mathcal{F}}\biggr[\mathcal{T}^{\mathcal{F}}\Bigr[e^{\int_{t_i}^{t_f}dt''\hat{\mathfrak{L}}_{\text{int}}(t'')}\Bigr]\hat{\rho}(t_i)\biggr]\biggr]
\end{split}
\end{equation}
We shall at first simplify the part corresponding to the time ordering of the gravitational field part. Considering $t_f-t_i$ to be a discrete composition of $N$ small time steps of time span $\Delta t$ and then taking the $\Delta t\rightarrow 0$ limit along with $N\rightarrow\infty$ limit, we obtain
\begin{equation}\label{2.19}
\begin{split}
\mathcal{T}^{\mathcal{F}}\Bigr[e^{\int_{t_i}^{t_f}dt'\hat{\mathfrak{L}}_{\text{int}}(t')}\Bigr]\hat{\rho}(t_i)&=\lim\limits_{\substack{{\Delta t\rightarrow 0}\\{N\rightarrow\infty}}}\exp\Bigr[\Delta t\sum\limits_{j=0}^N\hat{\mathfrak{L}}_{\text{int}}(t_j)+\frac{1}{2}(\Delta t)^2\sum\limits_{\substack{{j,k=0}\\{j>k}}}^N[\hat{\mathfrak{L}}_{\text{int}}(t_j),\hat{\mathfrak{L}}_{\text{int}}(t_k)]\Bigr]\hat{\rho}(t_i)\\
&=e^{\int_{t_i}^{t_f}dt \hat{\mathfrak{L}}_{\text{int}}(t)+\frac{1}{2}\int_{t_i}^{t_f}dt\int_{t_i}^{t_f}dt'\Theta(t-t')[\hat{\mathfrak{L}}_{\text{int}}(t),\hat{\mathfrak{L}}_{\text{int}}(t')]}\hat{\rho}(t_i)~.
\end{split}
\end{equation} 
\end{widetext}
In order to simplify the above result, one needs to compute the commutator bracket $[\hat{\mathfrak{L}}_{\text{int}}(t),\hat{\mathfrak{L}}_{\text{int}}(t')]\hat{\rho}(t_i)$. A little bit of algebra reveals that one can re-express the above relation as
\begin{equation}\label{2.20}
[\hat{\mathfrak{L}}_{\text{int}}(t),\hat{\mathfrak{L}}_{\text{int}}(t')]\hat{\rho}(t_i)=-[[\hat{H}_{\text{int}}(t),\hat{H}_{\text{int}}(t')],\hat{\rho}(t_i)].
\end{equation}
The commutation relation can be obtained between the interaction Hamiltonians from eq.(\ref{2.12}) up to the second order in the coupling constant $\kappa$ as (for a single mode of the BEC)
\begin{equation}\label{2.21}
\begin{split}
&[\hat{H}_{\text{int}}(t),\hat{H}_{\text{int}}(t')]\simeq\frac{4\kappa^2\gamma_\beta^2c_s^4}{V}\sum\limits_{\mathbf{k},s}\sum\limits_{\mathbf{k}',s'}[\hat{\delta h}^I_{\mathbf{k},s}(t)\varepsilon^s_{ij}(\mathbf{k}),\\&\hat{\delta h}^I_{\mathbf{k}',s'}(t')\varepsilon^{s'}_{lm}(\mathbf{k}')]\hat{k}_\beta^i\hat{k}_\beta^j\hat{k}_\beta^l\hat{k}_\beta^m|\hat{\psi}_{\mathbf{k}_\beta}(t)|^2|\hat{\psi}_{\mathbf{k}_\beta}(t')|^2~.
\end{split}
\end{equation}
From the above equation, we already find out that the commutator between the two interaction Hamiltonians is already quadratic in the two $\delta \hat{h}$ terms, and as a result, we can simply drop any noise fluctuation contribution from the $|\hat{\psi}_{\mathbf{k}_\beta}(t)|^2$ terms. 
The noise term can be identified as \cite{Super_Condensate_OTM,Super_Condensate_OTM_Lett}
\begin{equation}\label{2.22}
\begin{split}
\delta\hat{N}_{ij}(t)\equiv&\frac{2\kappa }{\sqrt{V}}\sum\limits_{s}\smashoperator{\sum\limits_{\substack{{\mathbf{k}}\\{|\mathbf{k}|\leq\Omega_m}}}}\delta \hat{h}^I_{\mathbf{k},s}(t)\epsilon^s_{ij}(\mathbf{k})~.
\end{split}
\end{equation}
Using eq.(s)(\ref{2.21},\ref{2.22}) the left hand side of eq.(\ref{2.20}) can be recast as
\begin{equation}\label{2.23}
\begin{split}
&[\hat{\mathfrak{L}}_{\text{int}}(t),\hat{\mathfrak{L}}_{\text{int}}(t')]\hat{\rho}(t_i)=-\gamma_\beta^2c_s^4[\delta\hat{N}_{ij}(t),\delta\hat{N}_{lm}(t')][\hat{k}_\beta^i\hat{k}_\beta^j\\&\hat{k}_\beta^l\hat{k}_\beta^m|\hat{\psi}_{\mathbf{k}_\beta}(t)|^2|\hat{\psi}_{\mathbf{k}_\beta}(t')|^2,\hat{\rho}(t_i)]~.
\end{split}
\end{equation}
One can now define two new operators given by
\begin{align}
\hat{\mathcal{K}}^{ij}_+(t)\hat{\rho}(t_i)&\equiv\hat{k}_\beta^i\hat{k}_\beta^j|\hat{\psi}_{\mathbf{k}_\beta}(t)|^2\hat{\rho}(t_i)~,\label{2.24}\\
\hat{\mathcal{K}}^{ij}_-(t)\hat{\rho}(t_i)&\equiv\hat{\rho}(t_i)\hat{k}_\beta^i\hat{k}_\beta^j|\hat{\psi}_{\mathbf{k}_\beta}(t)|^2~.\label{2.25}
\end{align}
\begin{widetext}
Using eq.(s)(\ref{2.24},\ref{2.25}), eq.(\ref{2.23}) can be recast
\begin{equation}\label{2.26}
\begin{split}
&[\hat{\mathfrak{L}}_{\text{int}}(t),\hat{\mathfrak{L}}_{\text{int}}(t')]\hat{\rho}(t_i)=-\gamma_\beta^2c_s^4[\delta\hat{N}_{ij}(t),\delta\hat{N}_{lm}(t')] \left(\hat{\mathcal{K}}^{ij}_+(t)\hat{\mathcal{K}}^{lm}_+(t')-\hat{\mathcal{K}}^{lm}_-(t')\hat{\mathcal{K}}^{ij}_-(t)\right)\hat{\rho}(t_i)~.
\end{split}
\end{equation}
It is now possible to recast eq.(\ref{2.18}) using eq.(s)(\ref{2.19},\ref{2.26}) as 
\begin{equation}\label{2.27}
\begin{split}
\hat{\rho}_{\text{BEC}}(t_f)=\mathcal{T}^{\text{BEC}}\biggr[e^{\int_{t_i}^{t_f}dt\hat{\mathfrak{L}}_0(t)-\frac{\gamma_\beta^2c_s^4}{2}[\delta\hat{N}_{ij}(t),\delta\hat{N}_{lm}(t')] \left(\hat{\mathcal{K}}^{ij}_+(t)\hat{\mathcal{K}}^{lm}_+(t')-\hat{\mathcal{K}}^{lm}_-(t')\hat{\mathcal{K}}^{ij}_-(t)\right)}\text{tr}_{\mathcal{F}}\Bigr[e^{\int_{t_i}^{t_f}dt\hat{\mathfrak{L}}_{\text{int}}(t)}\hat{\rho}(t_i)\Bigr]\biggr]~.
\end{split}
\end{equation}
\end{widetext}
One can consider that at $t=t_i$ the gravitational wave has started interacting with the BEC, and as a result one can consider the initial density matrix for the system to be a tensor product of the density matrix corresponding to the matter part and the field part as
\begin{equation}\label{2.28}
\hat{\rho}(t_i)=\hat{\rho}_{\text{BEC}}(t_i)\otimes\hat{\rho}_{\mathcal{F}}(t_i)~.
\end{equation}
One can now easily set the initial time to be $t_i=0$. The trace over the field can be assigned by a new quantity as
\begin{equation}\label{2.29}
\hat{\mathcal{W}}[\mathcal{K}_+,\mathcal{K}_-]\equiv\text{tr}_{\mathcal{F}}\Bigr[e^{\int_{t_i}^{t_f}dt\hat{\mathfrak{L}}_{\text{int}}(t)}\hat{\rho}(t_i)\Bigr]~.
\end{equation}
Making use of the decomposition in eq.(\ref{2.28}), eq.(\ref{2.29}) can be recast in the following form
\begin{equation}\label{2.30}
\begin{split}
\hat{\mathcal{W}}[\mathcal{K}_+,\mathcal{K}_-]&=\text{tr}_{\mathcal{F}}\Bigr[e^{\int_{t_i}^{t_f}dt\hat{\mathfrak{L}}_{\text{int}}(t)}\hat{\rho}_{\mathcal{F}}(t_i)\Bigr]\hat{\rho}_{\text{BEC}}(t_i)~,\\
\simeq \hat{\rho}_{\text{BEC}}(t_i)+&\frac{1}{2}\int_{t_i}^{t_f} dt\int_{t_i}^{t_f} dt'\left \langle\hat{\mathfrak{L}}_{\text{int}}(t)\hat{\mathfrak{L}}_{\text{int}}(t')\right\rangle_{\mathcal{F}}\hat{\rho}_{\text{BEC}}(t_i)
\end{split}
\end{equation}
where the higher order terms in the interaction Hamiltonian have been truncated. We shall later on combine the effects in an overall exponential term as the higher order terms are very small. The expectation value in eq.(\ref{2.30}) can be expressed as 
\begin{widetext}
\begin{equation}\label{2.31}
\begin{split}
\left \langle\hat{\mathfrak{L}}_{\text{int}}(t)\hat{\mathfrak{L}}_{\text{int}}(t')\right\rangle_{\mathcal{F}}\hat{\rho}_{\text{BEC}}(t_i)=&-\gamma_\beta^2c_s^4\Bigr[\langle \delta\hat{N}_{ij}(t)\delta\hat{N}_{lm}(t')\rangle_{\mathcal{F}}\hat{\mathcal{K}}^{ij}_+(t)\hat{\mathcal{K}}^{lm}_+(t')-\langle \delta\hat{N}_{lm}(t')\delta\hat{N}_{ij}(t)\rangle_{\mathcal{F}}\hat{\mathcal{K}}^{lm}_-(t')\hat{\mathcal{K}}^{ij}_+(t)\\
-&\langle \delta\hat{N}_{ij}(t)\delta\hat{N}_{lm}(t')\rangle_{\mathcal{F}}\hat{\mathcal{K}}^{ij}_-(t)\hat{\mathcal{K}}^{lm}_+(t')+\langle \delta\hat{N}_{lm}(t')\delta\hat{N}_{ij}(t)\rangle_{\mathcal{F}}\hat{\mathcal{K}}^{ij}_-(t)\hat{\mathcal{K}}^{lm}_-(t')\Bigr]\hat{\rho}_{\text{BEC}}(t_i)~.
\end{split}
\end{equation}
Using eq.(\ref{2.30}) in eq.(\ref{2.31}), and after some algebraic manipulations, one arrives at the following relation
\begin{equation}\label{2.32}
\begin{split}
\hat{\mathcal{W}}[\mathcal{K}_+,\mathcal{K}_-]&=\exp\biggr[-\frac{\gamma_\beta^2c_s^4}{2}\int_{t_i}^{t_f}dt\int_{t_i}^{t}dt'\left\langle\left\{\delta\hat{N}_{ij}(t),\delta\hat{N}_{lm}(t')\right\}\right\rangle_{\mathcal{F}}\Bigr[\hat{\mathcal{K}}^{ij}_+(t)\hat{\mathcal{K}}^{lm}_+(t')+\hat{\mathcal{K}}^{ij}_-(t)\hat{\mathcal{K}}^{lm}_-(t')-\hat{\mathcal{K}}^{ij}_-(t)\hat{\mathcal{K}}^{lm}_+(t')\\&-[\hat{\mathcal{K}}^{lm}_-(t')\hat{\mathcal{K}}^{ij}_+(t)\Bigr]-\frac{\gamma_\beta^2c_s^4}{2}\int_{t_i}^{t_f}dt\int_{t_i}^{t}dt'[\delta\hat{N}_{ij}(t),\delta\hat{N}_{lm}(t')]\Bigr(\hat{\mathcal{K}}^{lm}_-(t')\hat{\mathcal{K}}^{ij}_+(t)-\hat{\mathcal{K}}^{ij}_-(t)\hat{\mathcal{K}}^{lm}_+(t')\Bigr)\biggr]\hat{\rho}_{\text{BEC}}(t_i)~.
\end{split}
\end{equation}
We now can define two new operators as
\begin{align}
\hat{\mathcal{K}}^{ij}_{c}(t)\hat{\rho}&\equiv\left[\hat{k}^i_\beta\hat{k}^j_\beta|\hat{\psi}_{\hat{\mathbf{k}}_\beta}(t)|^2,\hat{\rho}\right]=\left(\hat{\mathcal{K}}^{ij}_{+}(t)-\hat{\mathcal{K}}^{ij}_{-}(t)\right)\hat{\rho}\label{2.33}\\
\hat{\mathcal{K}}^{ij}_{a}(t)\hat{\rho}&\equiv\left\{\hat{k}^i_\beta\hat{k}^j_\beta|\hat{\psi}_{\hat{\mathbf{k}}_\beta}(t)|^2,\hat{\rho}\right\}=\left(\hat{\mathcal{K}}^{ij}_{+}(t)+\hat{\mathcal{K}}^{ij}_{-}(t)\right)\hat{\rho}~.\label{2.34}
\end{align}
Using the above two relations, one can recast eq.(\ref{2.27}) as
\begin{equation}\label{2.35}
\begin{split}
\hat{\rho}_{\text{BEC}}(t_f)=&\mathcal{T}^{\text{BEC}}\biggr[\exp\left(\int_{t_i}^{t_f}dt\hat{\mathfrak{L}}_0(t)-\frac{\gamma_\beta^2c_s^4}{2}\int_{t_i}^{t_f}dt\int_{t_i}^{t}dt'[\delta\hat{N}_{ij}(t),\delta\hat{N}_{lm}(t')] \hat{\mathcal{K}}^{ij}_c(t)\hat{\mathcal{K}}^{lm}_a(t')\right)\\&\times\exp\left(-\frac{\gamma_\beta^2c_s^4}{2}\int_{t_i}^{t_f}dt\int_{t_i}^{t}dt'\langle\{\delta\hat{N}_{ij}(t),\delta\hat{N}_{lm}(t')\}\rangle_{\mathcal{F}} \hat{\mathcal{K}}^{ij}_c(t)\hat{\mathcal{K}}^{lm}_c(t')\right)\hat{\rho}_{\text{BEC}}(t_i)\biggr]\\
=&\mathcal{T}^{\text{BEC}}\biggr[\exp\left(\int_{t_i}^{t_f}dt\hat{\mathfrak{L}}_0(t)-\frac{\gamma_\beta^2c_s^4}{2}\int_{t_i}^{t_f}dt\int_{t_i}^{t}dt'[\delta\hat{N}_{ij}(t),\delta\hat{N}_{lm}(t')] \hat{\mathcal{K}}^{ij}_c(t)\hat{\mathcal{K}}^{lm}_a(t')\right)\hat{\tilde{\mathcal{W}}}(t_f,t_i)\hat{\rho}_{\text{BEC}}(t_i)\biggr]
\end{split}
\end{equation}
where we have defined a new operator as
\begin{equation}\label{2.36}
\begin{split}
\hat{\tilde{\mathcal{W}}}(t_f,t_i)\hat{\rho}\equiv\exp\left(-\frac{\gamma_\beta^2c_s^4}{2}\int_{t_i}^{t_f}dt\int_{t_i}^{t}dt'\langle\{\delta\hat{N}_{ij}(t),\delta\hat{N}_{lm}(t')\}\rangle_{\mathcal{F}} \hat{\mathcal{K}}^{ij}_c(t)\hat{\mathcal{K}}^{lm}_c(t')\right)\hat{\rho}~.
\end{split}
\end{equation}
\end{widetext}
This is one of the main analytical results in our paper.
 It is important to note that there is no expectation value taken for the commutator of the two noise parameters as the commutator comes out to be a number. One can in this case consider the mode expansion for the graviton mode operator using the mode function corresponding to the graviton state. For example, if one considers graviton states without any squeezing, then the noise commutator has the form
\begin{equation}\label{2.37}
[\delta\hat{N}_{ij}(t),\delta\hat{N}_{lm}(t')]=-i\zeta_{ijlm}(t,t')
\end{equation}
where $\zeta_{ijlm}(t,t')$ is a number. 

\noindent The analytical form of $\zeta_{ijlm}(t,t')$ reads
\begin{equation}\label{2.38}
\begin{split}
\zeta_{ijlm}(t,t')\equiv& \frac{2\kappa^2}{5\pi^2c^2}(\delta_{ij}\delta_{lm}+\delta_{im}\delta_{jl}-\frac{2}{3}\delta_{ij}\delta_{lm})\\
\times& \left(\frac{\sin(\Omega_m(t-t'))}{(t-t')^2}-\frac{\Omega_m\cos(\Omega_m(t-t'))}{t-t'}\right).
\end{split}
\end{equation}
We now start with a basic maximally entangled state for the BEC (at $t=0$)
\begin{equation}\label{2.39}
|\psi_{\text{BEC}}\rangle=\frac{1}{\sqrt{2}}\left(|\mathbf{k}_{\beta_1}\rangle\otimes|0_{\beta_2}\rangle+|0_{\beta_1}\rangle\otimes|\mathbf{k}_{\beta_2}\rangle\right)
\end{equation}
where $\hat{k}_\beta^i|\mathbf{k}_{\beta_a}\rangle=k_{\beta_a}^i|\mathbf{k}_{\beta_a}\rangle$ with $a=1,2$ and $i$ denoting the spatial index. Here, $|\mathbf{k}_{\beta_a}\rangle$ denotes the state corresponding to the $a$-th phonon mode of the BEC \footnote{For the rest of our calculation, $a$,  $\beta_a$ in the subscript of $k$ denotes both 1 and 2 modes of the phonon.}. The primary idea behind using the eigenstates of the transverse wave-number operator in the Fourier space is that the BEC is characterized by superposing matter waves corresponding to each phonon where the condensate forms in the Fourier space. In order to truly capture the decoherence effect due to interacting gravitons in a BEC one needs to look at such eigenstates (corresponding to individual phonon mode frequencies) instead of position states. Considering the effect of the environmental quantum gravitational field on the BEC, one can write down the initial state of the BEC-graviton system as (at $t=t_i$)
\begin{equation}\label{2.39a}
\begin{split}
|\psi_i\rangle&=\frac{1}{\sqrt{2}}\left(|\mathbf{k}_{\beta_1}\rangle\otimes|0_{\beta_2}\rangle+|0_{\beta_1}\rangle\otimes|\mathbf{k}_{\beta_2}\rangle\right)\otimes |h\rangle\\
&=|\psi_{\text{BEC}}\rangle\otimes|h\rangle
\end{split}
\end{equation} 
with $|h\rangle$ denoting the initial graviton state. It is important to note from the above equation that the BEC-graviton states are separable in nature. Here, we impose the following normalization conditions
\begin{equation}\label{2.40a}
\langle \mathbf{k}_{\beta_a}| \mathbf{k}_{\beta_a}\rangle=1,~\langle 0_{\beta_a}| 0_{\beta_a}\rangle \text{ and }\langle h| h\rangle=1~.
\end{equation}
In principle, one can set $t_i=0$ and
obtain the initial density matrix of the condensate system as
\begin{equation}\label{2.40}
\begin{split}
\hat{\rho}_{\text{BEC}}(0)&=\text{tr}_{\mathcal{F}}\left[|\psi_i\rangle\langle\psi_i|\right]=\hat{\rho}_{11}+\hat{\rho}_{12}+\hat{\rho}_{21}+\hat{\rho}_{22}
\end{split}
\end{equation} 
where the components of the reduced density matrix read
\begin{align}
\hat{\rho}_{11}&=\frac{1}{2}\left(|\mathbf{k}_{\beta_1}\rangle\otimes|0_{\beta_2}\rangle\right)\left(\langle \mathbf{k}_{\beta_1}|\otimes\langle 0_{\beta_2}|\right)\label{2.41a}\\
\hat{\rho}_{12}&=\frac{1}{2}\left(|\mathbf{k}_{\beta_1}\rangle\otimes|0_{\beta_2}\rangle\right)\left(\langle 0_{\beta_1}|\otimes\langle \mathbf{k}_{\beta_2}|\right)\label{2.41b}\\
\hat{\rho}_{21}&=\frac{1}{2}\left(|0_{\beta_1}\rangle\otimes|\mathbf{k}_{\beta_2}\rangle\right)\left(\langle \mathbf{k}_{\beta_1}|\otimes\langle 0_{\beta_2}|\right)\label{2.41c}\\
\hat{\rho}_{22}&=\frac{1}{2}\left(|0_{\beta_1}\rangle\otimes|\mathbf{k}_{\beta_2}\rangle\right)\left(\langle 0_{\beta_1}|\otimes\langle \mathbf{k}_{\beta_2}|\right)\label{2.41d}~.
\end{align}
Here, it is very important to see that we can still represent the reduced density matrix in eq.(\ref{2.40}) as $\hat{\rho}_{\text{BEC}}(0)=|\psi_{\text{BEC}}\rangle\langle \psi_{\text{BEC}}|$.
From section (\ref{S4}), we shall see that the two different paths will be represented by two coherent atom laser beams that are entangled at the source where the source represents a BEC with a weak-coupling constant inside of a harmonic trap potential. Now gravitons couple with one of the beams, due to Bremsstrahlung, the quantum state of the graviton changes. The state of the entire BEC system coupled to the environmental gravitons after a time $t_f$ will read
\begin{equation}\label{2.41e}
\begin{split}
|\psi_f\rangle&=\frac{1}{\sqrt{2}}|\mathbf{k}_{\beta_1},0_{\beta_2}\rangle\otimes |h,\mathbf{k}_{\beta_1}\rangle+\frac{1}{\sqrt{2}}|0_{\beta_1},\mathbf{k}_{\beta_2}\rangle\otimes|h,\mathbf{k}_{\beta_2}\rangle
\end{split}
\end{equation}
where $|k_{\beta_1},k_{\beta_2}\rangle\equiv |k_{\beta_1}\rangle\otimes|k_{\beta_2}\rangle$. The coupling of the BEC with the gravitons can be easily inferred from the form of the interaction Hamiltonian in eq.(\ref{2.12}). We can see that in eq.(\ref{2.41e}), we have kept the state of the BEC to be the same as the gravitational back-reaction on the BEC is negligible. 
In order to  investigate the effect of the Bremsstrahlung of gravitons, one needs to look at the reduced density matrix given as
\begin{equation}\label{2.41f}
\begin{split}
\hat{\rho}_{\text{BEC}}(t_f)&=\hat{\rho}_{11}+\hat{\rho}_{22}+e^{i\Delta(t_f)}\hat{\rho}_{12}+e^{-i\Delta^*(t_f)}\hat{\rho}_{21}
\end{split}
\end{equation}
where $\Delta(t_f)$ is the influence functional given by 
\begin{equation}\label{2.41g}
e^{i\Delta(t_f)}\equiv  \langle h,\mathbf{k}_{\beta_2}|h,\mathbf{k}_{\beta_1}\rangle~.
\end{equation}
The influence phase has a real as well as imaginary part where the imaginary part is called the decoherence function. The decoherence function leads to the entanglement loss between the two beams of the atom lasers. Our primary aim is to calculate the decoherence functional. Acting with the $\hat{\tilde{\mathcal{W}}}(t_f,0)$ operator, defined in eq.(\ref{2.36}), on the initial BEC state we obtain
\begin{equation}\label{2.41}
\begin{split}
\hat{\tilde{\mathcal{W}}}(t_f,0)\hat{\rho}_{\text{BEC}}(0)=\hat{\rho}_{11}+\hat{\rho}_{22}+\hat{\rho}_{12}e^{-\Gamma(t_f)}+\hat{\rho}_{21}e^{-\Gamma(t_f)}
\end{split}
\end{equation}
where in the above equation, the analytical form of $\Gamma(t_f)$ reads
\begin{equation}\label{2.42}
\begin{split}
\Gamma(t_f)&=\frac{\gamma_\beta^2c_s^4}{2}\int_0^{t_f}dt\int_{0}^{t}dt'\langle\{\delta\hat{N}_{ij}(t),\delta\hat{N}_{lm}(t')\}\rangle_{\mathcal{F}}\\&\times\Delta\left(k_\beta^ik_\beta^j|\psi_{\mathbf{k}_\beta}(t)|^2\right)\Delta\left(k_\beta^lk_\beta^m|\psi_{\mathbf{k}_\beta}(t')|^2\right)~.
\end{split}
\end{equation}
Note that $\hat{\rho}_{11}$ and $\hat{\rho}_{22}$ remains unaltered. In eq.(\ref{2.42}), $\Delta\left[k_\beta^ik_\beta^j|\psi_{\mathbf{k}_\beta}(t)|^2\right]$ is defined as
\begin{equation}\label{2.43}
\begin{split}
\Delta\left[k_\beta^ik_\beta^j|\psi_{\mathbf{k}_\beta}(t)|^2\right]\equiv k_{\beta_1}^ik_{\beta_1}^j|\psi_{\mathbf{k}_{\beta_1}}(t)|^2-k_{\beta_2}^ik_{\beta_2}^j|\psi_{\mathbf{k}_{\beta_2}}(t)|^2
\end{split}
\end{equation}
with the form of $\psi_{\mathbf{k}_{\beta_i}}$ for $i=1,2$ being given by \cite{PhononBEC3,ThesisMatthew,Super_Condensate_OTM,Super_Condensate_OTM_Lett} $\psi_{\mathbf{k}_{\beta_i}}\simeq\alpha_{\beta_i}e^{-i\omega_{\beta_i}t}+\beta_{\beta_i}e^{i\omega_{\beta_i}t}$ such that all noise contributions to the time-dependent pseudo-Goldstone bosons are dropped. Using the symmetry property of the anti-commutator, one can recast eq.(\ref{2.42}) as
\begin{widetext}
\begin{equation}\label{2.44}
\begin{split}
\Gamma(t_f)&=\frac{\gamma_\beta^2c_s^4}{4}\int_0^{t_f}dt\int_{0}^{t_f}dt'\langle\{\delta\hat{N}_{ij}(t),\delta\hat{N}_{lm}(t')\}\rangle_{\mathcal{F}}\Delta\left(k_\beta^ik_\beta^j|\psi_{\mathbf{k}_\beta}(t)|^2\right)\Delta\left(k_\beta^lk_\beta^m|\psi_{\mathbf{k}_\beta}(t')|^2\right)~.
\end{split}
\end{equation}
We shall now operate with the exponential term containing the noise-noise commutator from eq.(\ref{2.35}) on $\hat{\tilde{\mathcal{W}}}(t_f,0)\hat{\rho}_{\text{BEC}}(0)$. This gives the following relation
\begin{equation}\label{2.45}
\begin{split}
&e^{i\frac{\gamma_\beta^2c_s^4}{2}\int_{t_i}^{t_f}dt\int_{t_i}^{t}dt'\zeta_{ijlm}\hat{\mathcal{K}}^{ij}_c(t)\hat{\mathcal{K}}^{lm}_a(t')}\hat{\tilde{\mathcal{W}}}(t_f,t_i)\hat{\rho}_{\text{BEC}}(t_i)=\hat{\rho}_{11}+\hat{\rho}_{22}+e^{-\Gamma(t_f)+i\Lambda(t_f)}\hat{\rho}_{12}+e^{-\Gamma(t_f)-i\Lambda(t_f)}\hat{\rho}_{21}
\end{split}
\end{equation}
where the analytical form of $\Lambda(t_f)$ is given by
\begin{equation}\label{2.46}
\begin{split}
\Lambda(t_f)&=\frac{\gamma_\beta^2c_s^4}{2}\int_0^{t_f}dt\int_{0}^{t}dt'\zeta_{ijlm}\Delta\left(k_\beta^ik_\beta^j|\psi_{\mathbf{k}_\beta}(t)|^2\right)\left(k_{\beta_1}^lk_{\beta_1}^m|\psi_{\mathbf{k}_{\beta_1}}(t')|^2+k_{\beta_2}^lk_{\beta_2}^m|\psi_{\mathbf{k}_{\beta_2}}(t')|^2\right)~.
\end{split}
\end{equation}.
\end{widetext}
It is now possible to write down the density matrix of the BEC, after the graviton interaction has occurred in the time interval $t_f-t_i=t_f$, as
\begin{equation}\label{2.47}
\begin{split}
\hat{\tilde{\rho}}_{\text{BEC}}(t_f)=\hat{\rho}_{11}+\hat{\rho}_{22}+e^{i\Delta(t_f)}\hat{\rho}_{12}+e^{-i\Delta^*(t_f)}\hat{\rho}_{21}
\end{split}
\end{equation}
 where $\Delta(t_f)\equiv \Lambda(t_f)+i\Gamma(t_f)$. 
 
\begin{figure}
\begin{center}
\includegraphics[scale=0.25]{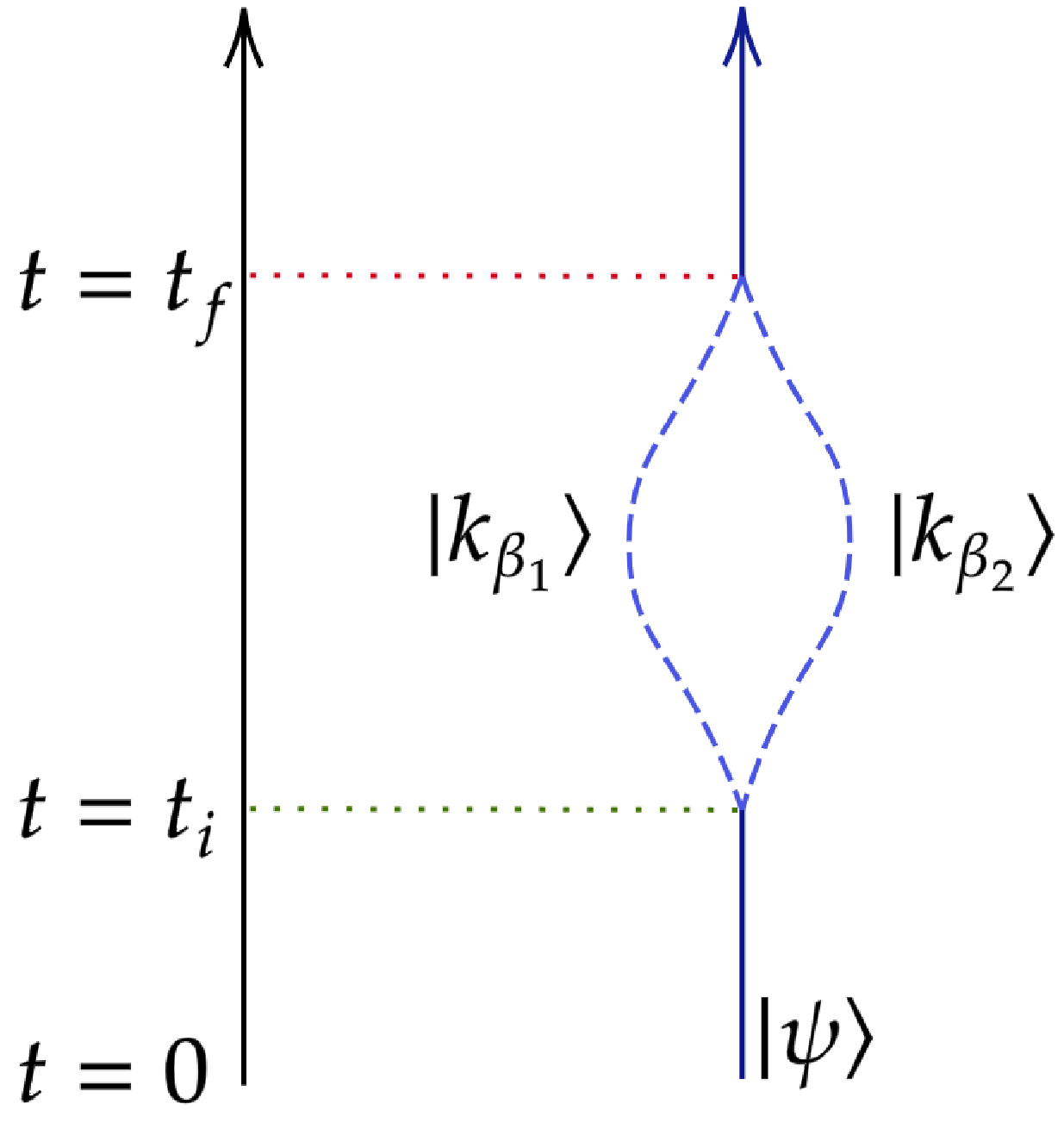}
 \caption{The superposition state of the two coherent BEC sources is presented in the diagram. For $t_i=0$, the state goes into an instant superposition state just after the formation of the BEC.\label{Superposition1}}
\end{center}
\end{figure}
\begin{figure}
\begin{center}
\includegraphics[scale=0.34]{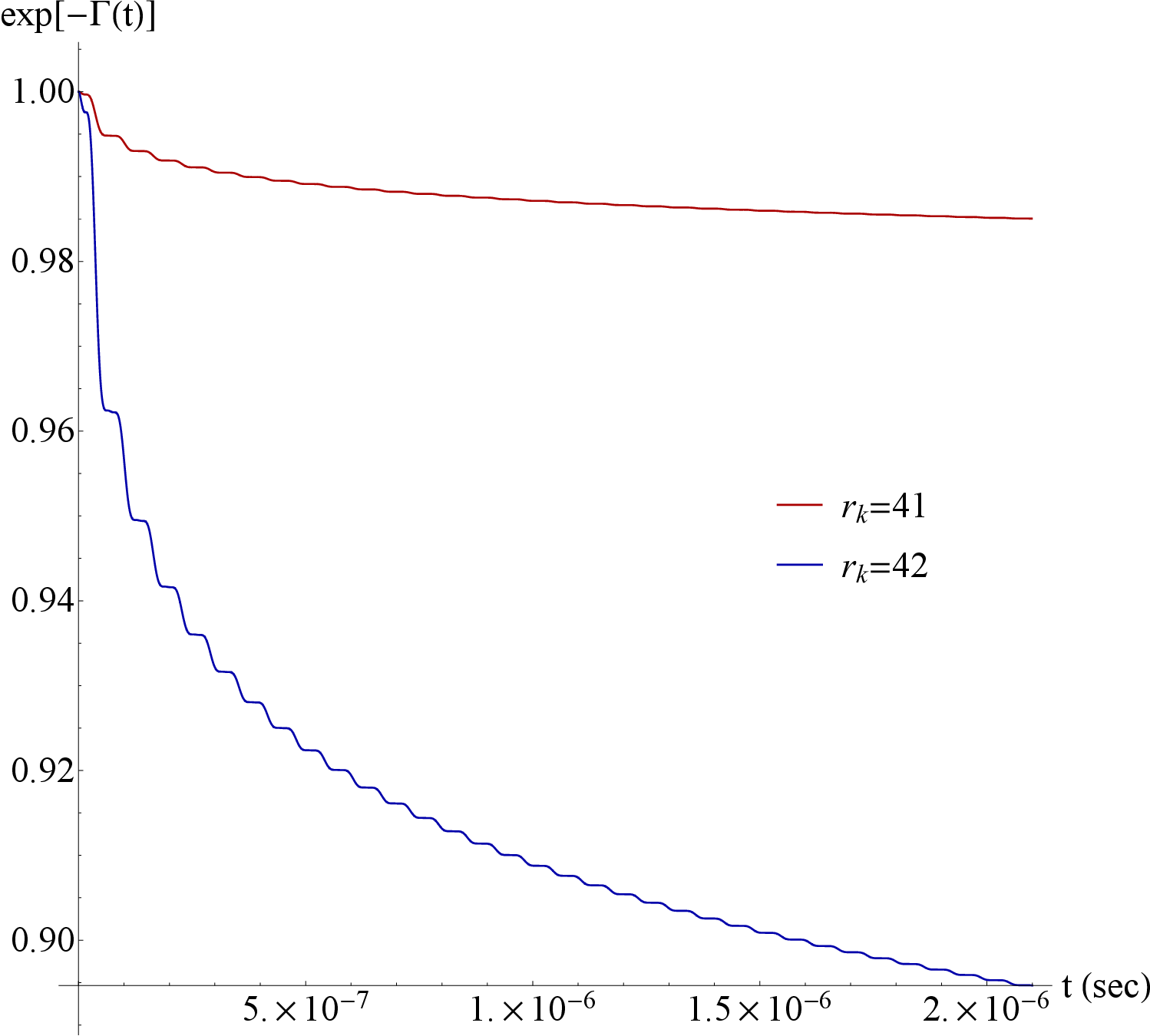}
\caption{We observe that time dependence of the decoherence term $e^{-\Gamma(t)}$ for change in time. We plot for squeezing parameter values $r_k=41,42$. We observe higher decoherence over time with a higher value of the graviton-squeezing parameter. \label{Decoherence0}}
\end{center}
\end{figure}

\noindent Eq.(\ref{2.47}) is identical to the reduced density matrix in eq.(\ref{2.41f}) and is a very important result in our paper. We see that the decaying exponential term depends on the two-point correlator of the noise term (or the anti-commutator) whereas the phase term is purely dependent upon the commutator between the same.
It is therefore evident that even if the two-point correlator is absent, the phase term will be present which after a very short time makes it impossible to write down $\hat{\rho}$ as $|\tilde{\psi}\rangle\langle\tilde{\psi}|$. This effect is generated purely due to the interaction of the gravitons (which are bosons as well) with the BEC. We call this BEC-graviton entangled system as a ``\textit{Bose-Einstein supercondensate}". We are not interested in the action of the Liouvile super-operator on the base Hamiltonian and therefore, we will safely propagate with the reduced density matrix in eq.(\ref{2.47}) as it captures the entire effect of the graviton interaction over the $t_f$ time interval. 
In Fig.(\ref{Superposition1}), we present a schematic diagram of the superposition of the two single-mode phonon states corresponding to two coherent BEC sources. From eq.(\ref{2.47}), it is easy to interpret that $\Gamma(t_f)$ is the decoherence rate of the superposed BEC states and one needs a continuous generation of BEC. Before going into experimental intricacies, we need to thoroughly investigate the form of the decoherence rate from eq.(\ref{2.44}). For the two-point correlator we shall make use of squeezed graviton states as with a zero squeezing value the general case can be obtained for which the commutator relation is given in eq.(\ref{2.38}). The value of the two-point correlator $\left\langle \{\delta \hat{N}_{ij}(t),\delta \hat{N}_{lm}(t')\}\right\rangle_{\mathcal{F}}$ from eq.(\ref{2.44}) takes the form \cite{Super_Condensate_OTM,Super_Condensate_OTM_Lett} \footnote{To obtain the form of the two point correlator one needs to make use of a few analytical relations. At first we need the summation condition for the polarization tensors given by $\sum_s\epsilon^{s*}_{ij}(\textbf{k})\epsilon^{s}_{lm}(\textbf{k})=\frac{1}{2}\left(\mathcal{P}_{il}\mathcal{P}_{jm}+\mathcal{P}_{im}\mathcal{P}_{jl}-\mathcal{P}_{ij}\mathcal{P}_{lm}\right)$ where the analytical form of the projection tensors read $\mathcal{P}_{ij}=\delta_{ij}-\frac{k_ik_j}{k^2}$. One also needs to make use of the angular integrals given by, $\int d\Omega=4\pi$, $\int d\Omega \frac{k^ik^j}{k^2}=\frac{4\pi}{3}\delta^{ij}$, and $\int d\Omega \frac{k^ik^jk^lk^m}{k^4}=\frac{4\pi}{15}\left(\delta^{ij}\delta^{lm}+\delta^{il}\delta^{jm}+\delta^{im}\delta^{jl}\right)$.}
 \begin{widetext}
 \begin{equation}\label{2.48}
\begin{split}
&\left\langle \{\delta \hat{N}_{ij}(t),\delta \hat{N}_{lm}(t')\}\right\rangle_{\mathcal{F}}\\&=\frac{2\kappa^2}{5\pi^2c^2}\left(\delta_{il}\delta_{jm}+\delta_{im}\delta_{jl}-\frac{2}{3}\delta_{ij}\delta_{lm}\right)\int_0^{\Omega_m}dk ~k\left(\cos\left(k(t-t')\right)\cosh 2r_k-\cos\left(k(t+t')-\phi_k\right)\sinh 2r_k\right)\\&=\frac{2\kappa^2}{5\pi^2c^2}\left(\delta_{il}\delta_{jm}+\delta_{im}\delta_{jl}-\frac{2}{3}\delta_{ij}\delta_{lm}\right)\Biggr(\cosh 2r_k\left[\frac{-1+\cos(\Omega_m (t-t'))+\Omega_m(t-t')\sin(\Omega_m (t-t'))}{(t-t')^2}\right]\\&+\sinh 2r_k\left[\frac{\cos\varphi_k+\cos(\Omega_m (t+t')-\phi_k)+\Omega_m(t+t')\sin(\Omega_m (t+t')-\varphi_k)}{(t+t')^2}\right]\Biggr)~.
\end{split} 
\end{equation}  
The above equation can be further recast as
\begin{equation}\label{eqredefined}
\begin{split}
\left\langle\{\delta\hat{N}_{ij}(t),\delta\hat{N}_{lm}(t')\}\right\rangle_{\mathcal{F}}&=\left(\delta_{il}\delta_{jm}+\delta_{im}\delta_{jl}-\frac{2}{3}\delta_{ij}\delta_{lm}\right)\langle\{\delta \hat{N}(t),\delta \hat{N}(t')\}\rangle_{\mathcal{F}}
\end{split}
\end{equation}
where the quantity $\langle\{\delta \hat{N}(t),\delta \hat{N}(t')\}\rangle_{\mathcal{F}}$ is defined as
\begin{equation}\label{eqredefined2}
\begin{split}
\langle\{\delta \hat{N}(t),\delta \hat{N}(t')\}\rangle_{\mathcal{F}}&\equiv\frac{2\kappa^2}{5\pi^2c^2}\Biggr(\cosh 2r_k\left[\frac{-1+\cos(\Omega_m (t-t'))+\Omega_m(t-t')\sin(\Omega_m (t-t'))}{(t-t')^2}\right]\\+&\sinh 2r_k\left[\frac{\cos\varphi_k+\cos(\Omega_m (t+t')-\phi_k)+\Omega_m(t+t')\sin(\Omega_m (t+t')-\varphi_k)}{(t+t')^2}\right]\Biggr) ~.
\end{split}
\end{equation}
\end{widetext}
We now assume that the two transverse wave numbers corresponding to BEC with fixed mode frequencies are shifted by a frequency factor that is proportional to the frequency of the incoming gravitational wave. As can be seen from Fig.(\ref{Superposition1}), the superposition is evident in the time interval $t_f-t_i$ and the two paths get displaced by a frequency value. 
The resonance happens at a frequency $\omega_\beta=\Omega/2$ with $\Omega$ being the frequency of the incoming gravitational wave. We consider that the change in frequency $\Delta\omega_\beta=|\omega_{\beta_2}-\omega_{\beta_1}|\leq\frac{\Omega}{2}$ which is equal to the effective frequency of the BEC. Again, we know that $k_\beta\simeq \frac{\omega_\beta}{c_s}$ \cite{PhononBEC3,PhononBEC4}. Hence, it is quite straightforward to assume $k_{\beta_1}=\frac{\omega_\beta}{c_s}-\frac{\Omega}{4c_s}$  and $k_{\beta_2}=\frac{\omega_\beta}{c_s}+\frac{\Omega}{4c_s}$. Here we consider the frequency gap to be the maximum considering the maximum separation of the paths in the Fourier space represented by Fig.(\ref{Superposition1}). Hence, the $\Delta\left(k_\beta^ik_\beta^j|\psi_{\mathbf{k}_\beta}(t)|^2\right)$ terms present in the decoherence rate reads
\begin{equation}\label{2.49}
\begin{split}
&\Delta\left(k_\beta^ik_\beta^j|\psi_{\mathbf{k}_\beta}(t)|^2\right)\\&=\frac{1}{c_s^2}\omega_{\beta_1}^i\omega_{\beta_1}^j\left(\alpha_{\beta_1}^2+\beta_{\beta_1}^2+2\alpha_{\beta_1}\beta_{\beta_1}\cos(2\omega_{\beta_1}t)\right)\\&-\frac{1}{c_s^2}\omega_{\beta_2}^i\omega_{\beta_2}^j\left(\alpha_{\beta_2}^2+\beta_{\beta_2}^2+2\alpha_{\beta_2}\beta_{\beta_2}\cos(2\omega_{\beta_2}t)\right)
\end{split}
\end{equation}
where in the argument of the sinusoidal functions, we have replaced the values of $\omega_{\beta_1}$ and $\omega_{\beta_2}$. 
The simplest model is obtained by taking either $\alpha_{\beta_a}$ (for $a=1,2$) or $\beta_{\beta_a}$ to be equal to zero. In our case, we neglect $\beta_{\beta_a}=0$\footnote{This choice can be supported by the form of the Bogoliubov coefficients as obtained in \cite{Super_Condensate_OTM,Super_Condensate_OTM_Lett}. For a choice of the classical part of the gravitational wave $h^{\text{cl}}(t,0)=\varepsilon e^{-t^2/\tau^2}\sin(\Omega t)$, $\hat{\alpha}_\beta=1+\mathcal{O}(\hat{N})$ and $\hat{\beta}_\beta=\mathcal{O}(\varepsilon)+\mathcal{O}(\hat{N})$ with $\varepsilon$ being the amplitude of the gravitational wave. As the decoherence parameter $\Gamma(t_f)\sim\mathcal{O}(N^2)$, one can safely ignore any $\mathcal{O}(\varepsilon,N)$ contributions from the Bogoliubov coefficients. This choice leaves one with $\alpha_\beta\simeq1$ and $\beta_\beta\sim 0$. (In \cite{Super_Condensate_OTM}, $h^{\text{cl}}(t,0)$ is given by the expression $\varepsilon e^{-t^2/\tau^2}\cos(\Omega t)$ which is a typographical error and the cosine function should be replaced by the sine function.) }. This is identical to the unperturbed form of the time-dependent part of the pseudo-Goldstone boson in \cite{PhononBEC3,PhononBEC4,
ThesisMatthew,Super_Condensate_OTM,
Super_Condensate_OTM_Lett}. This will take away the time dependence of the difference term in the left-hand side of eq.(\ref{2.49}). Note that to generalize the result but at the same time keep the calculations straightforward, one can simply set $\alpha_{\beta_a}=\beta_{\beta_a}$ and $\alpha_{\beta_1}\simeq\alpha_{\beta_2}=\alpha_{\beta}$\footnote{The assumption $\alpha_\beta\simeq\beta_\beta$ is valid only if the initial solution of the non-interacting BEC wave function is considered to be $\psi_{\mathbf{k}_\beta^{(0)}}=\mathcal{a}e^{-i\omega_{\beta}t}+\mathcal{b}e^{i\omega_{\beta}t}$ with non-vanishing value of $\mathcal{b}$. The $\mathcal{b}$ is generally taken to be zero \cite{PhononBEC3,Super_Condensate_OTM,Super_Condensate_OTM_Lett} for getting rid of the negative energy modes.}. We shall start with the simplest possible case for which the difference term from eq.(\ref{2.49}) becomes time-independent ($\beta_{\beta_a}\sim0$). It is also quite natural to set the remaining Bogoliubov coefficient to unity as $\alpha_\beta,\beta_\beta\sim 1$. The form of eq.(\ref{2.42}) (one can also work with eq.(\ref{2.44})) takes the form
\begin{equation}\label{2.50}
\begin{split}
\Gamma(t_f)=\frac{8\gamma_\beta^2\omega_\beta^2\Omega^2}{3}\int_0^{t_f}dt\int_{0}^tdt'\langle \{\delta\hat{N}(t),\delta\hat{N}(t')\}\rangle_{\mathcal{F}}.
\end{split}
\end{equation}
To obtain the above result, we have made use of the result that $\mathbf{k}_{\beta_1}\cdot\mathbf{k}_{\beta_2}=k_{\beta_1}k_{\beta_2}\cos \theta_{\beta}=\frac{\omega_{\beta_1}\omega_{\beta_2}}{c_s^2}$ where the angle between the two transverse wave-number vectors has been taken to be zero. 
Squeezed graviton states can be present in primordial gravitational waves from the inflationary time \cite{KannoSodaTokuda,KannoSodaTokuda2}. We consider the incoming gravitational wave frequency to be $\Omega=1$ Hz and as a result, the resonance frequency of the BEC will lie at about $\omega_\beta=0.5$ Hz. It is important to note that one should correctly estimate the coupling constant of the BEC. In general, BEC can be attained for weakly as well as strongly interacting bosonic systems. For a BEC with weakly interacting bosons \cite{LinWolfe}, one can consider the coupling constant to be quite small $\lambda\sim 10^{-7}\ll1$. Before obtaining the decoherence effects, we shall investigate the amplitude of the effect induced by graviton noise. The two point correlator in eq.(\ref{eqredefined2}), is of the order of $\langle \{\delta\hat{N}(t),\delta\hat{N}(t')\}\rangle\sim \frac{\hbar G}{c^5 }e^{2r_k}f(t,t')$ with high graviton squeezing and $f(t,t')$ carrying the time dependence and a dimension of $[f]=T^{-2}$.  We can see that $\langle \{\delta\hat{N}(t),\delta\hat{N}(t')\}\rangle\sim10^{-65}f(t,t')$ even before executing the time integrals. Hence, the pre-factor in eq.(\ref{2.50}), $\frac{8\omega_\beta^2\Omega^2\gamma_\beta^2}{3}$, must be very large for $\Gamma$ being in a measurable range. It is important to note that the $\gamma_\beta$ term in the pre-factor is not a BEC-graviton coupling constant and purely dependent on the condensate parameters by the relation (as has been given earlier after eq.(\ref{2.3a})) $\gamma_\beta=\frac{L^3_\beta}{\lambda c}(\frac{m^2 c^2}{\hbar^2})$ when $\tilde{\sigma}\sim \frac{mc^2}{\hbar}$. Hence, there is no restriction or bound on the value of $\gamma_\beta$ and for our system, we shall estimate the value of $\gamma_\beta$. As we are considering $L_\beta=10^{-3}$ m, $\lambda\sim 10^{-7}$, and the effective mass of the Bose-Einstein condensate to be $m\sim10^{-25}$ kg, we observe the value of $\gamma_\beta$ to be $\gamma_\beta\sim 10^{24}-10^{25}$ sec. Hence the pre-factor $\frac{8\omega_\beta^2\Omega^2\gamma_\beta^2}{3}$ is of the order of $\frac{8\omega_\beta^2\Omega^2\gamma_\beta^2}{3}\sim 10^{48}-10^{50}$ $\text{sec}^{-2}$ for the frequency choices $\omega_\beta=0.5$ and $\Omega=1$ Hz. Another important thing to note that this $\gamma_\beta$ factor is inversely related to $\lambda$ and therefore for a strongly coupled BEC the $\gamma_\beta$ factor is lower when all other parameters are kept same. Hence, the rate of decay is lower for a strongly coupled BEC system and this phenomenon is controlled by this factor $\gamma_\beta$. For the origin of this factor $\gamma_\beta$, please refer to our first work \cite{Super_Condensate_OTM}. It is important to note that the analytical form of $\gamma_\beta$ gets fixed while writing down the total action for the BEC-graviton system in eq.(\ref{2.4}) \cite{Super_Condensate_OTM}. Considering $r_k\sim 24$, and executing the time integrals in eq.(\ref{2.50}), the order of magnitude for the decoherence factor is obtained to be of the order of $\Gamma(t_f)\sim 10^{-16}$ for $t_f\sim 2~\mu\text{s}$. Hence, the amplitude corresponding to the off-diagonal elements of the reduced density matrix has the value $1- 10^{-16}$, which is indeed very small but quite highly enhanced due to the effect induced by the graviton squeezing. In the absence of any graviton squeezing the amplitude corresponding to the off-diagonal terms of the density matrix takes the value of $1- 10^{-37}$. In \cite{PhononBEC2} a classical gravitational wave detection scenario using a Bose-Einstein condensate was proposed where the amplitude was found to be $1-N\varepsilon$ where $N$ denotes the number of atoms in the condensate and $\varepsilon\sim 10^{-21}$. In \cite{PhononBEC2}, the state of the condensate was considered to be a coherent state. As discussed in \cite{PhononBEC2}, the enhancement by $N$ cannot truly reduce the gap of twenty orders of magnitude. Later, in \cite{PhononBEC3,PhononBEC4}, enhancement of the sensitivity of BEC towards classical gravitational wave was proposed via the inclusion of squeezing of the phonon modes and implementation of parametric resonance. In our case, we have two squeezing parameters to enhance the sensitivity of the BEC highly, one is the squeezing of the graviton states and the other one is the squeezing of the phonon modes\footnote{For a detailed discussion, please refer to \cite{Super_Condensate_OTM} where quantum metrological techniques have been used to write down the quantum gravitational Fisher information and from there we have obtained the sensitivity of the BEC as a graviton detector. As the quantum gravitational Fisher information is inversely related to the uncertainty in the amplitude of the gravitational wave, the standard quantum Fisher information and its quantum gravity-induced part are treated on an equal footing. In this case, we see that both the phonon squeezing and the graviton squeezing increase the sensitivity of the BEC towards the gravitons from incoming primordial gravitational waves whereas for low phonon squeezing and no graviton squeezing the BEC is not sensitive to the gravitons from incoming primordial gravitational waves.}. If the phonon squeezing (as will be discussed later in subsection (\ref{Sub1})) is of the order of $r_\omega\sim 8.5-9.0$, then the amplitude term has a numerical value $1-10^{-2}$ to $1-10^{-1}$ when the final time of measurement is of the range $t_f\sim 1-2 ~\mu\text{s}$. This implementation of phonon squeezing, hugely compensates the gap observed earlier in \cite{PhononBEC2}. One can also just rely on the squeezing alone of the gravitons where with a squeezing value of $r_k\sim 41-42$ one can achieve such enhancements. It is evident that for gravitons with no squeezing, the decoherence rate will be very small as a result the decoherence will be minimal. We, therefore, consider the decoherence due to squeezed gravitons and for different squeezing parameters with a fixed squeezing angle ($\varphi_k=\frac{\pi}{2}$ used here) and plot it against time in Fig.(\ref{Decoherence0}). We observe from Fig.(\ref{Decoherence0}) that with a higher value of the graviton squeezing, the decoherence is higher in the entangled BEC system. Another important aspect is that the exponential decay factor has a wave-like decay pattern which can be an effect of the random fluctuations from the gravitons. For the graviton-squeezing of $r_k=42$, we observe that a $10\%$ coherence loss happens over a time period of $1-2~\mu\text{s}$ whereas for $r_k=41$, it takes almost $4\times10^7$ sec for observing a similar coherence loss. This result, hence, signifies that detecting graviton signatures with lower squeezing values will be very difficult. Such a squeezing value can only prevail in primordial gravitational waves where the gravitons are squeezed at the time of inflation \cite{KannoSodaTokuda}.
A Bose-Einstein condensate with a high enough coupling constant can be also formed in this case but detecting the effects of graviton-induced decoherence will be much more difficult in such a scenario. An estimation of the coupling constant can be found using the analysis in \cite{PhononBEC3}. A gross estimate of the coupling constant reads \cite{PhononBEC3,ThesisMatthew}
\begin{equation}\label{2.51}
\begin{split}
\lambda\sim \frac{m^4c_s^2c^3}{\rho\hbar^3}
\end{split}
\end{equation}
where $m$ is the mass of the atoms and $\rho=T_{00}$ being the energy density of the BEC with $T_{\mu\nu}=\frac{1}{\sqrt{-g}}\frac{\delta S}{\delta g^{\mu\nu}}$. The action $S$ is the simple action corresponding to the BEC and classical gravity interaction. The graviton part in the action can be neglected in this analysis as it will not contribute towards any significant changes in the energy density of the system. The energy density $\rho$ can be estimated from the analysis in \cite{PhononBEC3} and the approximate value of the coupling constant $\lambda$ comes out to be $\lambda\sim 10^{11}$. In such a strongly coupled BEC, even with a graviton squeezing $r_k\sim 76$, there is only a $2\times10^{-5}\%$ loss of coherence over a time interval $t_f-t_i\sim 10^{4}$ sec. Hence, for truly detecting graviton signatures, one needs to deal with weakly coupled BEC systems. 

\noindent Next, we consider a strongly coupled BEC system with coupling parameter $\lambda=10$. In general, the decoherence time is determined by the time by which the exponential factor decays to $1/e$ of its initial value or $\Gamma(t_f)\sim 1$. As, we have discussed in the earlier analysis, for a strongly coupled BEC system, the graviton squeezing must be very high for the detection of significant decoherence over a smaller period in the BEC-graviton system. 
\begin{figure}[ht!]
\begin{center}
\includegraphics[scale=0.34]{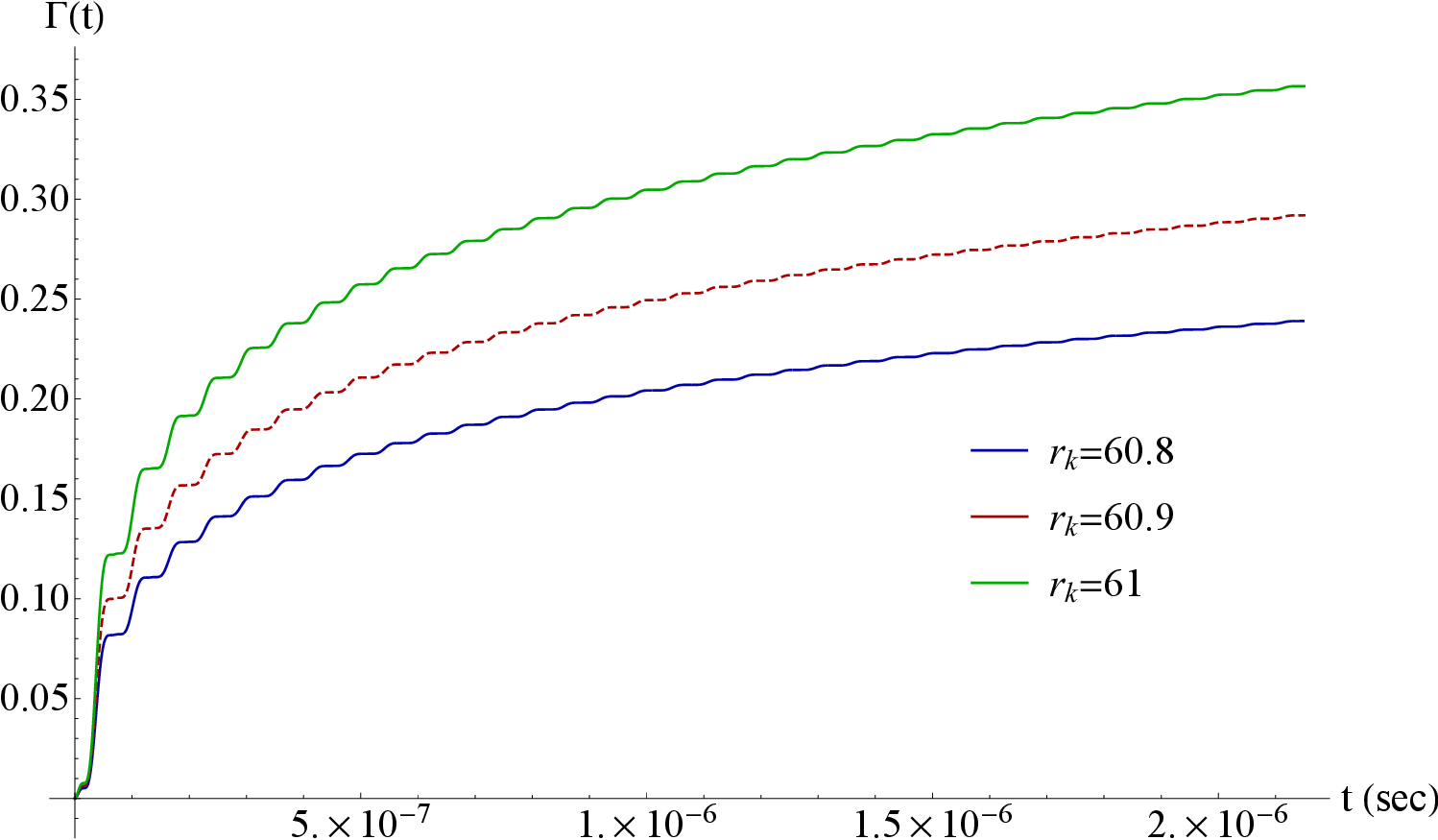}
\caption{We plot the decoherence function $\Gamma(t)$ against the observation time $t$ for different values of the graviton squeezing. We find out that for $r_k=60.8$, the decoherence time is at around $t\sim 29.275$ sec which decreases drastically even with a very small increase in the graviton squeezing. \label{Decoherence_Time}}
\end{center}
\end{figure} 
From Fig.(\ref{Decoherence_Time}), we find out that the decoherence time is very small for gravitons with higher squeezing. With $r_k=60.8$, we find the decoherence time to be $t\simeq 29.275$ sec whereas these times are $t\simeq0.585$ sec for $r_k=60.9$, and $t\simeq0.024$ sec for $r_k=61$. In Fig.(\ref{Decoherence_Time}), we have plotted upto $t=2 \mu\text{S}$ as it properly depicts the behaviour of the decoherence factor $\Gamma$. Now the squeezing needed for a sensible measurement of decoherence time due to Bremsstrahlung can be decreased drastically for BEC with lower coupling constant values and we consider the coupling constant value $\lambda=10^{-7}$ as has been considered earlier.

\begin{figure}[ht!]
\begin{center}
\includegraphics[scale=0.34]{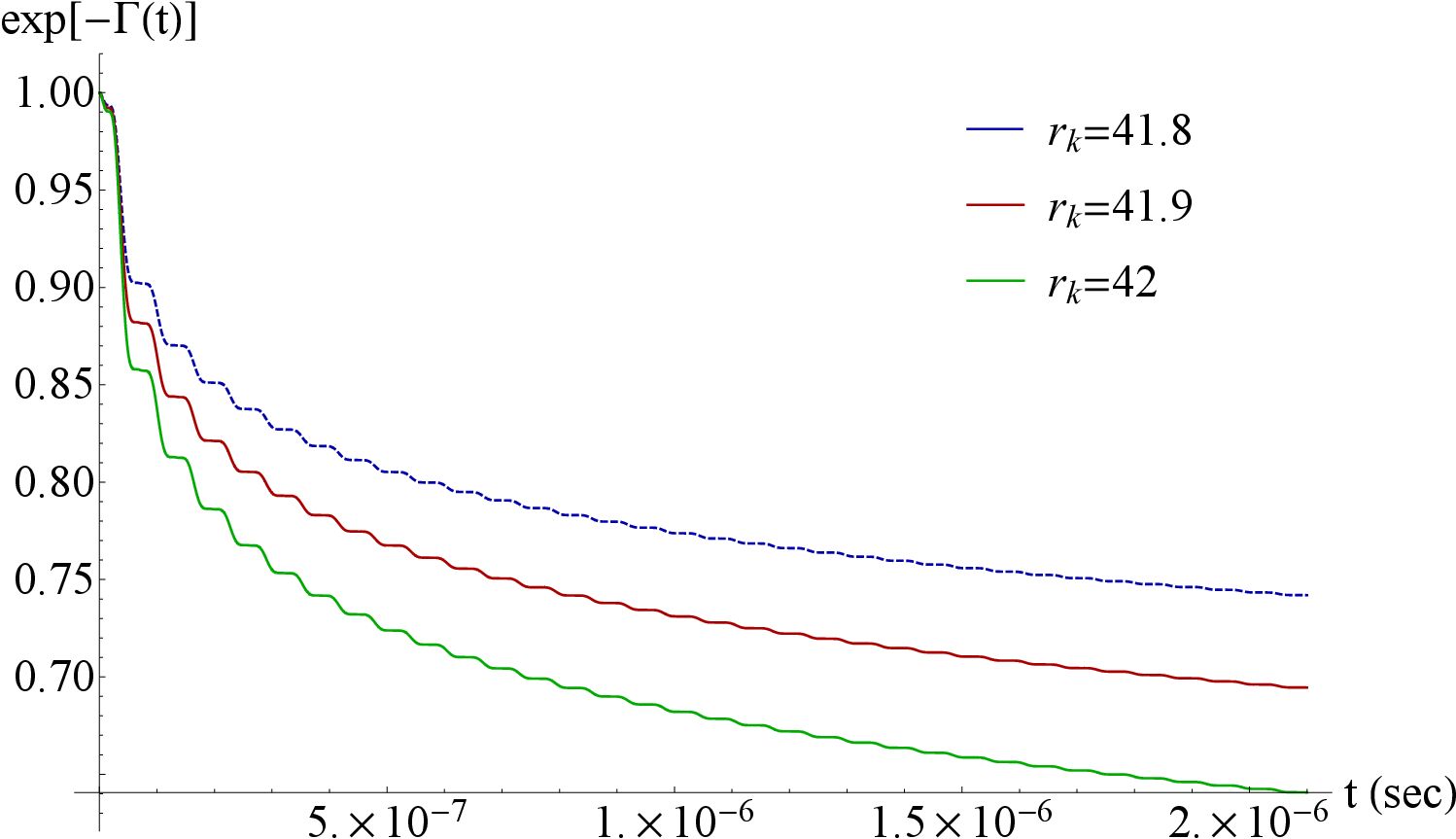}
\caption{We observe the time dependence of the decoherence term $e^{-\Gamma(t)}$ with change in time. We observe a faster coherence-loss over the same $2$ $\mu$s time interval for a different form factor of the time-dependent part of the pseudo-Goldstone boson compared to Fig.(\ref{Decoherence0}).
\label{Decoherence1}}
\end{center}
\end{figure}

\noindent Finally, we carry out the same analysis for a slightly non-trivial structure of the time-dependent part of the pseudo-Goldstone bosons. For primordial gravitational waves and for $t_f<1$ sec, $\omega_\beta t\pm\frac{\Omega t}{4}<1$ (when $\Omega\sim 1$ Hz and $t\ll 1$ sec). For such primordial gravitational waves, we can use a Taylor series expansion for the cosine functions in eq.(\ref{2.54}) when $\alpha_{\beta_a}=\beta_{\beta_a}$ and $\alpha_{\beta_a},\beta_{\beta_a}\sim1$. Using the same parameters as used in Fig.(\ref{Decoherence0}), we observe a faster decay of the decoherence factor over the same time period of 2 $\mu$s time-interval which has been plotted in Fig.(\ref{Decoherence1}) (with graviton squeezing $r_k=41.8,41.9$ and $42$). The decoherence time is observed to be ($\exp[-\Gamma(t_d)]=1/e$) $t_d=1.25$ ms for $r_k=42$.
\subsection{Can squeezing the phonons help in reducing the decoherence time?}\label{Sub1}
\noindent In this subsection, we shall do a back-of-the-envelope calculation showing the effect of the squeezing of the phonons of the BEC on the decoherence time due to Bremsstrahlung from the BEC \textit{supercondensate}. It is important to note that the transverse wave number $k_\beta$ is directly related to the number of phonons $n_\beta$ of the BEC via the relation $k_\beta=\frac{\pi n_\beta}{L_{\beta}}$, with $L_\beta$ denoting the length of the side of the cubic box inside which the BEC is created (generally this box is experimentally mimicked using a harmonic trap potential). The transverse wave number operator can be expressed as $\hat{k}_\beta=\frac{\pi \hat{n}_\beta}{L_\beta}$. It is quite straightforward to understand that $\hat{n}_{\beta}$ gives the number operator which in terms of the creation and annihilation operator takes the form $\hat{n}_\beta=\hat{a}^\dagger_\beta\hat{a}_\beta$ with $[\hat{a}_\beta,\hat{a}^\dagger_\beta]=1$. Under the action of a squeezing operator $\hat{S}({z_\omega})=e^{\frac{1}{2}\left(z_\omega^*\hat{a}_\beta^2-z_\omega\hat{a}^{\dagger2}_\beta\right)}$, we want to get the eigenvalue $k_\beta$ corresponding to a state $|\mathbf{k}_\beta\rangle$. Here, $z_\omega=r_\omega e^{i\phi_\omega}$ with $r_\omega$ being the squeezing parameter and $\phi_\omega$ being the squeezing angle, and $\hat{S}(z_\omega)\hat{S}^\dagger(z_\omega)=\hat{\mathbb{1}}$. Under the action of the squeezing operator, the annihilation and creation operator changes as
\begin{equation}\label{sq.1}
\begin{split}
\hat{S}^\dagger(z_\omega)\hat{a}_\beta\hat{S}(z_\omega)&=\hat{a}_\beta\cosh(r_\omega)-e^{i\phi_\omega}\sinh(r_\omega)\hat{a}^\dagger_\beta\\
\hat{S}^\dagger(z_\omega)\hat{a}_\beta^\dagger\hat{S}(z_\omega)&=\hat{a}^\dagger_\beta\cosh(r_\omega)-e^{-i\phi_\omega}\sinh(r_\omega)\hat{a}_\beta~.
\end{split}
\end{equation}
The action of the transverse-wave number operator on a single squeezed state of the BEC then takes the form
\begin{equation}\label{sq.2}
\begin{split}
\hat{k}_\beta|k_\beta^{\text{sq.}}\rangle&=\frac{\pi\hat{a}^\dagger_\beta\hat{a}_\beta}{L_\beta}\hat{S}(z_\omega)|k_\beta\rangle\\
&=\frac{\pi\hat{S}(z_\omega)\hat{S}^\dagger(z_\omega)\hat{a}^\dagger_\beta\hat{S}(z_\omega)\hat{S}^\dagger(z_\omega)\hat{a}_\beta}{L_\beta}\hat{S}(z_\omega)|k_\beta\rangle
\end{split}
\end{equation} 
where we have made use of the property of the squeezing operator, namely, $\hat{S}(z_\omega)\hat{S}^\dagger(z_\omega)=\hat{\mathbb{1}}$. Making use of eq.(\ref{sq.1}), one can recast the above equation as
\begin{equation}\label{sq.3}
\begin{split}
\hat{k}_\beta|k_\beta^{\text{sq.}}\rangle&=\frac{\pi\hat{S}(z_\omega)}{L_\beta}\Bigr(\hat{n}_\beta\cosh^2(r_\omega)+(\hat{n}_\beta+1)\sinh^2(r_\omega)\\&-\frac{\sinh(2r_\omega)}{2}\left(e^{i\phi_\omega}\hat{a}^{\dagger2}_\beta+e^{-i\phi_\omega}\hat{a}_\beta^2\right)\Bigr)|k_\beta\rangle~.
\end{split}
\end{equation}
Even for a weakly coupled BEC, $n_\beta$ is quite large ($n_\beta\gg1$) and as a result we can write down the action of the annihilation operator on the BEC state simply as $\hat{a}_\beta|k_\beta\rangle=\sqrt{n_\beta}\left|\frac{\pi(n_\beta-1)}{L_\beta}\right>\simeq \sqrt{n_\beta}|k_\beta\rangle$ and the action of the creation operator as $\hat{a}_\beta^\dagger|k_\beta\simeq \sqrt{n}_\beta|k_\beta$. This helps us to recast eq.(\ref{sq.3}) as
\begin{equation}\label{sq.4}
\begin{split}
\hat{k}_\beta|k^{\text{sq.}}_\beta\rangle&\simeq \frac{\pi \hat{S}(z_\omega)}{L_\beta}n_\beta\left[ \cosh(2r_\omega)-\sinh(2r_\omega)\cos\phi_\omega\right]|k_\beta\rangle.
\end{split}
\end{equation}
For a phonon-squeezing angle $\phi_\omega=\pi/2$\footnote{Squeezing the phonons at specific angles has been done previously and experimentally quite achievable \cite{Chelkowski,Johnsson}.}, the above equation takes the form
\begin{equation}\label{sq.5}
\begin{split}
\hat{k}_\beta|k^{\text{sq.}}_\beta\rangle&=\frac{\pi\cosh (2r_\omega)n_\beta}{L_\beta}\hat{S}(z_\omega)|k_\beta\rangle\\
&=\cosh(2r_\omega)k_\beta|k_\beta^{\text{sq.}}\rangle~.
\end{split}
\end{equation}
Now $\omega_\beta=c_sk_\beta$, and as a result under phonon squeezing $\omega_\beta$ goes to $\cosh(2r_\omega)\omega_\beta$. We can directly see the effect of phonon squeezing simply by carefully observing the form of the decoherence factor $\Gamma(t_f)$ in eq.(\ref{2.50}). Under the effect of phonon squeezing, $\Gamma(t_f)$ takes the form
\begin{equation}\label{sq.6}
\begin{split}
\Gamma(t_f)^{\text{sq.}}&\simeq \frac{8}{3}\gamma_\beta^2\cosh^2(2r_\omega)\omega_\beta^2\\&\times\int_0^{t_f}dt\int_0^tdt'\langle \{\delta\hat{N}(t),\delta\hat{N}(t')\}\rangle_\mathcal{F}\\
&=\cosh^2(2r_\omega)\Gamma(t_f).
\end{split}
\end{equation}
Researchers have achieved a phonon-squeezing of $7.2$ dB which is equivalent to a squeezing with squeezing parameter $r_\omega=0.83$ \cite{GuLiWuYang}.  In \cite{PhononBEC3}, it was shown that the semi-classical BEC-gravitational wave model supports a maximum squeezing of $r_\omega\approx27$. We find out that for a proper measurement of the decoherence effect, one needs a sufficiently faster decay of the decoherence term $\exp[-\Gamma(t_f)]$. To fulfil such a condition one needs at least a graviton squeezing higher than $r_k=40$. We aim to use the phonon squeezing as a control parameter such that it gives the experimentalists a wider control over the graviton-detection scenario. From eq.(\ref{sq.6}), it is straightforward to argue that the decoherence term now takes the form $\exp[-\cosh^2(2r_\omega)]\Gamma(t_f)]$. Similarly, when time dependence is present in eq.(\ref{2.49}), we also find quite an identical behaviour. We consider the graviton squeezing to be $r_k=\{19.8,20\}$ with phonon squeezing to be $r_\omega=\{11.0,11.2\}$. 
\begin{figure}[ht!]
\begin{center}
\includegraphics[scale=0.336]{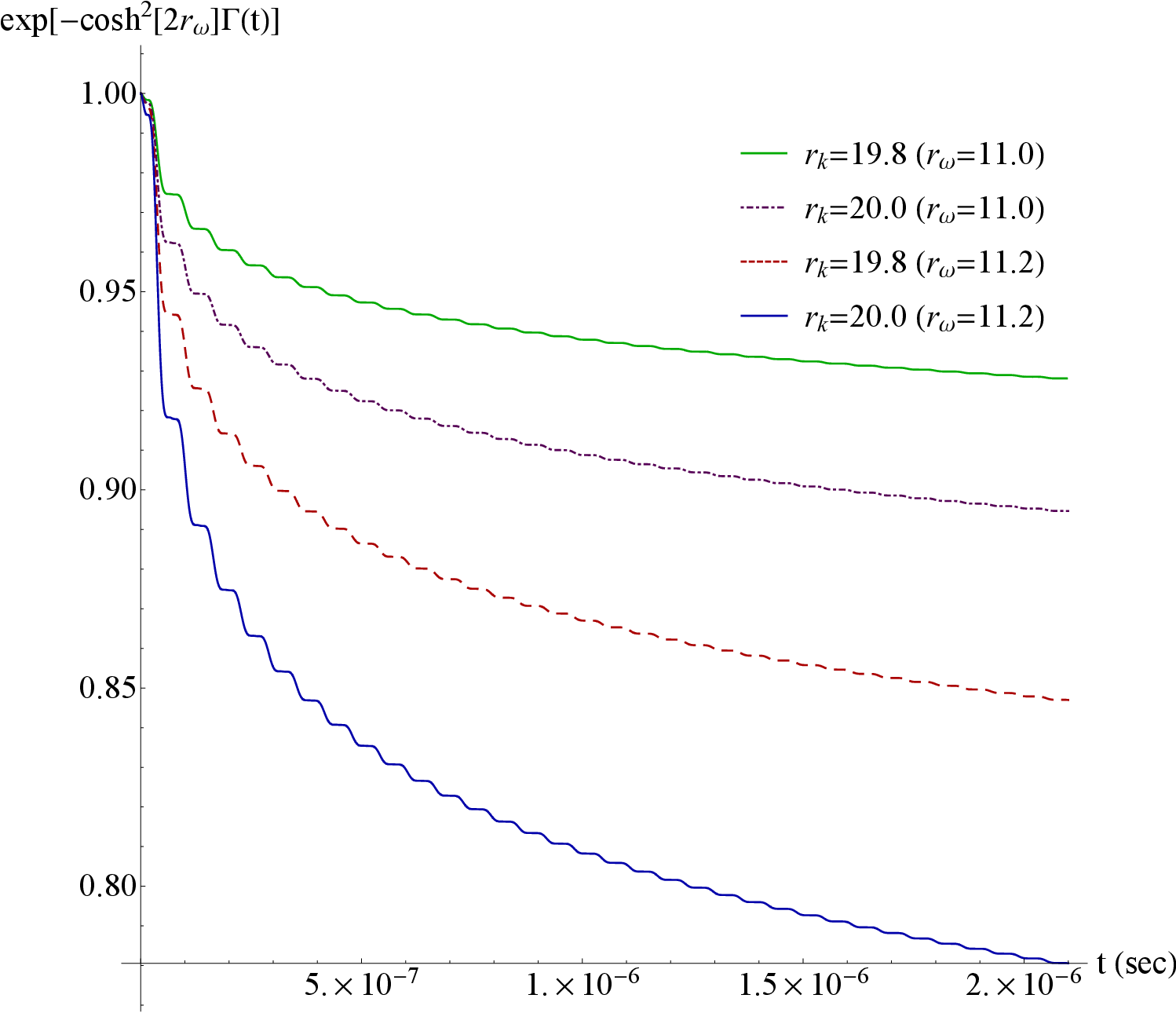}
\caption{We observe the time dependence of the decoherence term $e^{-\cosh^2(2r_\omega)\Gamma(t)}$ corresponding to a BEC with phonons squeezed by a squeezing parameter $r_\omega$ with change in time. We plot for graviton-squeezing parameter values $r_k=19.8,20$ and phonon-squeezing parameter values $r_k=11.0,11.2$. We observe higher decoherence over time with a higher value of the phonon-squeezing parameter. \label{Decoherencef}}
\end{center}
\end{figure}
We observe from Fig.(\ref{Decoherencef}) that even for a graviton-squeezing as low as $r_k\sim 20$, there is a $10\%$ coherence loss over a $2~\mu$s time period when the phonon squeezing is $r_\omega\sim11.0$. The most important observation is that just by increasing the phonon-squeezing parameter by $\Delta r=0.2$, the coherence loss gets to $20\%$ over the same time period for the $r_k=20.0$ case. This behaviour depicts that via tuning the squeezing of the phonons, one has a high chance of detecting gravitons that are coming with a lower graviton-squeezing. Now $r=11.0$ is equivalent to a phonon-squeezing of $-10\log_{10}(-2r_\omega)=95.54$ dB. It is extremely difficult to create a squeezing of the order of $100$ dB experimentally but we strongly believe that such squeezing can be achieved by means of a decades time. For a phonon-squeezing parameter as high as $r_\omega=1.0$, a $10\%$ coherence loss is observed over a $2~\mu$s time period when the graviton-squeezing parameter is $r_k=40$.
\subsection{Entanglement degradation due to Bremsstrahlung from the BEC \textit{supercondensate}}
\begin{figure}[ht!]
\begin{center}
\includegraphics[scale=0.24]{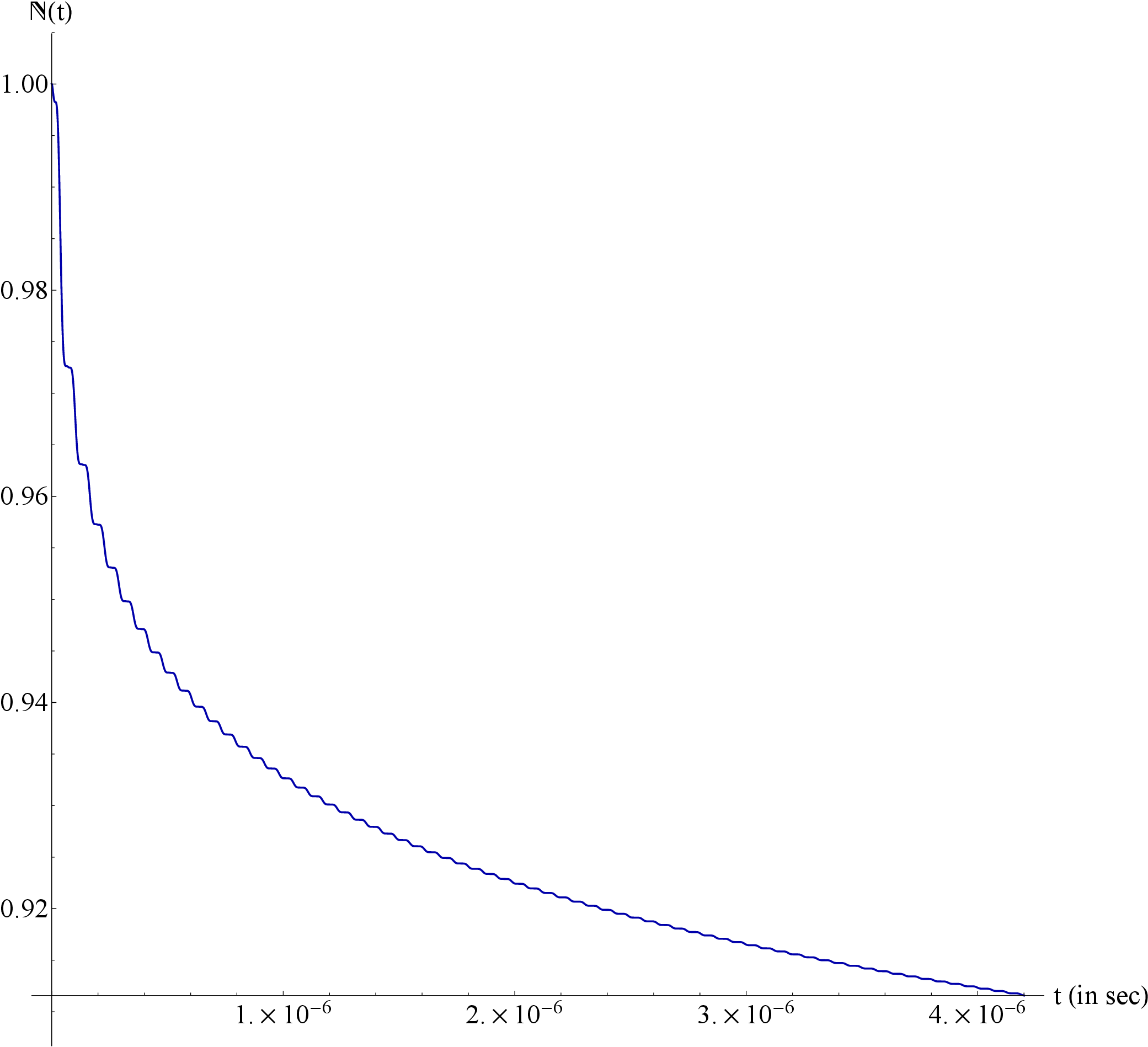}
\caption{We plot the logarithmic negativity of the system against time. We observe that the logarithmic negativity follows a jiggly decay pattern amplifying the fact that the entanglement loss is due to the noise induced by gravitons.\label{Negativity}}
\end{center}
\end{figure}

\noindent We shall now focus on the time evolution and the evolution of entanglement over time for the initial maximally entangled density matrix with the state defined in eq.(\ref{2.39}). We shall make use of the entanglement negativity to inspect the status of the density matrix over time. Taking the partial trace of the density matrix in eq.(\ref{2.47}), we obtain
\begin{equation}\label{2.52}
\hat{\tilde{\rho}}^T_{\text{BEC}}(t_f)=\frac{1}{2}\begin{pmatrix}
0&&0&&0&&e^{i\Delta(t_f)}\\
0&&1&&0&&0\\
0&&0&&1&&0\\
e^{-i\Delta^*(t_f)}&&0&&0&&0
\end{pmatrix}~.
\end{equation}
The eigenvalues corresponding to the above matrix read
\begin{equation}\label{2.53}
\lambda_1=\lambda_2=\frac{1}{2},~\lambda_3=\frac{1}{2}e^{-\Gamma(t_f)}, \text{ and }\lambda_4=-\frac{1}{2}e^{-\Gamma(t_f)}~.
\end{equation}
The entanglement negativity is defined as \cite{FuentesMann}
\begin{equation}\label{2.54}
\begin{split}
\mathbb{N}\equiv\log_2||\hat{\rho}^T||=\log_2\Bigr[1+\sum\limits_{\lambda_-}\left(\lambda_--|\lambda_-|\right)\Bigr]
\end{split}
\end{equation}
where $\lambda_-$ denotes the negative eigenvalues corresponding to the partial transpose of the density matrix $\hat{\rho}$. The entanglement negativity corresponding to the density matrix in eq.(\ref{2.39}) takes the form
\begin{equation}
\begin{split}
\mathbb{N}(t_f)=\log_{2}\left[1+e^{-\Gamma(t_f)}\right]~.
\end{split}
\end{equation}
Using the same analytical result of $\Gamma(t_f)$ from eq.(\ref{2.50}), one can plot the change in the logarithmic negativity over time which is depicted in Fig.(\ref{Negativity}).
We observe that the logarithmic negativity of the system degrades over time. The jiggly decay pattern is similar to Fig.(\ref{Decoherence0}) and confirms the entanglement loss via the emission of bremsstrahlung due to the noise induced by gravitons. The pattern also shows several time steps where the degradation rate becomes suddenly slower and again it amplifies. In the next section, we shall propose an experimental set up that shall help us to detect the signatures of gravitons by measuring the loss of coherence due to such Bremsstrahlung, induced by the noise of gravitons.
\section{Using atom interferometry with ultra-cold atom lasers to detect graviton signature}\label{S4}
\noindent After the creation of a Bose-Einstein condensation in a laboratory \cite{1Nobel2001,2Nobel2001}, physicists at Massachusetts Institute of Technology devised a way to create atom-lasers from a trapped Bose-Einstein condensate of Sodium atoms \cite{3Nobel2001}. The primary reason for using atom lasers is that they produce a coherent set of propagating atomic waves where the atom lasers are created from a BEC inside of a harmonic trap. Therefore, if one can construct two such coherent beams and via using atom interferometry \cite{AtomInterferometry,AtomInterferometry2} check for interference between them, any kind of decoherence in the interference pattern should be theoretically measurable. Before proceeding further, it is important to note that we need two such BEC sources that are maximally entangled, as can be seen from eq.(\ref{2.39}). The first step is to construct a multi-beam (preferably a two-beam) atom laser where each beam is coherent with each other. In \cite{DualbeamAtomLaser}, a two-beam atom laser was produced by outcoupling oppositely polarized components of an all-optical BEC. Another novel technique of producing coherent atom beam splitting from a single far-detuned laser was experimentally observed and reported in \cite{MultibeamAtomLaser}. 
Making use of a magnetic trap potential and Bragg diffraction from two optical standing wave gratings one can produce a continuous atom laser coherently split into multiple momentum states (in \cite{MultibeamAtomLaser} three such atom laser beams were observed). These momentum states have slightly different momentum values as required by our analysis as well. The only difference in our work is that the base frequency is considered to be the same while calculating the decoherence rate which was later altered by one-fourth of the incoming gravitational wave frequency. This can be properly adjusted in an experimental scenario. Now, the second step is to entangle the separated beam of coherent atom lasers which will mimic the initial state of the system as in eq.(\ref{2.39}). Very recently in \cite{NatureEntangled}, a matter wave interferometry has been done between two entangled matter waves inside a high-fineness cavity. Making use of direct collisional interactions \cite{NatureEntangled2,NatureEntangled3,
NatureEntangled4,NatureEntangled5,
ScienceEntangled,ScienceEntangled2,ScienceEntangled3}  or Coulomb interactions \cite{NatureEntangled6,PRLEntangled} along with relative atom number squeezing between matter wave in spatially separated traps \cite{NatureEntangled2,NatureEntangled4,ScienceEntangled3}, one can experimentally entangles two atoms. Such entanglement has also been observed between atoms via mapping of internal entanglement onto the relative atom number in different momentum eigenstates \cite{PRLEntangled2}. In \cite{NatureEntangled}, the entanglement between the external momentum states of different atoms was attained using a cavity
quantum-electrodynamical set up. This technique though relies on the strong collective coupling between the atoms and the optical cavity. Quantum momentum kicks are provided to the atom via the use of two-photon Raman transitions which results in the splitting and recombination of the matter wave packet. In this experimental set up, an entanglement of $18.5$ dB was achieved. The injected two-atom beams are passed through a Mach-Zehnder typer atom interferometer to check whether the output data is below the standard quantum limit. Using atom interferometers for classical gravitational wave detection has already been proposed in \cite{PRLGravWave,PRAGravWave}. Very recently in \cite{BoseGraviton} decoherence rate in a matter wave interferometer due to electromagnetic interactions has been considered. This result has then been used in the quantum gravity-induced entanglement of masses protocol or the QGEM protocol. Based on the experimental advancements discussed up to this, we propose a graviton detector using ultra-cold atom lasers from a weakly coupled BEC in Fig.(\ref{Graviton_Detector_final}). One can see from Fig.(\ref{Graviton_Detector_final}) that our entire set up is placed inside a cavity and can be thought of as a combination of two Mach-Zender type atom-interferometers. We now give a step-by-step description of our proposed BEC-based graviton detector in Fig.(\ref{Graviton_Detector_final}).

\begin{enumerate}
\item\textcolor{blue}{\textit{Continuous generation of an ultra-cold atom laser}}

\noindent It is important to understand that we are proposing to detect graviton signatures due to primordial gravitational waves as they will come with inherent graviton squeezing (from the inflationary period). As a result, it is impossible to know when the gravitational wave passes through Earth. Hence, one needs a continuous generation of BEC inside a harmonic trap. Recently in \cite{ContinuousBEC}, continuous-Bose-Einstein condensate was produced using a continuous-wave condensate made of strontium atoms which can last indefinitely. In \cite{ContinuousBEC}, from a steady-state narrow-line magneto-optical trap ${}^{84}\text{Sr}$ atoms were outcoupled continuously to a guide and then were loaded into a crossed beam-dipole trap forming a larger reservoir with a small dimple. The dimple then started to get heavily populated by atoms creating a steady state BEC. This process is quite ground-breaking and we propose to use this method for generating a continuous BEC state and from there generate a continuously flowing atom laser as can be seen from Fig.(\ref{Graviton_Detector_final})\footnote{Our theoretical model is based on a weakly coupled BEC with coupling parameter $\lambda\sim10^{-7}$. 
We hope that such weakly coupled BEC can be formed which can primarily be found in Bose gases.}. It is to be noted that our theoretical model is based on a pure BEC which is almost impossible to create in a laboratory as it requires laser cooling to the absolute zero temperature. It is still quite feasible to use BEC with a temperature couple of nano-Kelvins above the absolute zero which shall also mimic quite nicely our current theoretical framework. Making use of the techniques in \cite{1Nobel2001,2Nobel2001,3Nobel2001}, one can use external energy sources (for example a periodic radio frequency pulse) to stimulate the BEC such that a chunk of ultra-cold atoms get released from the trap producing an ultra-cold atom laser accelerating freely under the effect of gravity.  

\begin{widetext}
\begin{center}
\begin{figure}[ht!]
\centering
\includegraphics[scale=0.335]{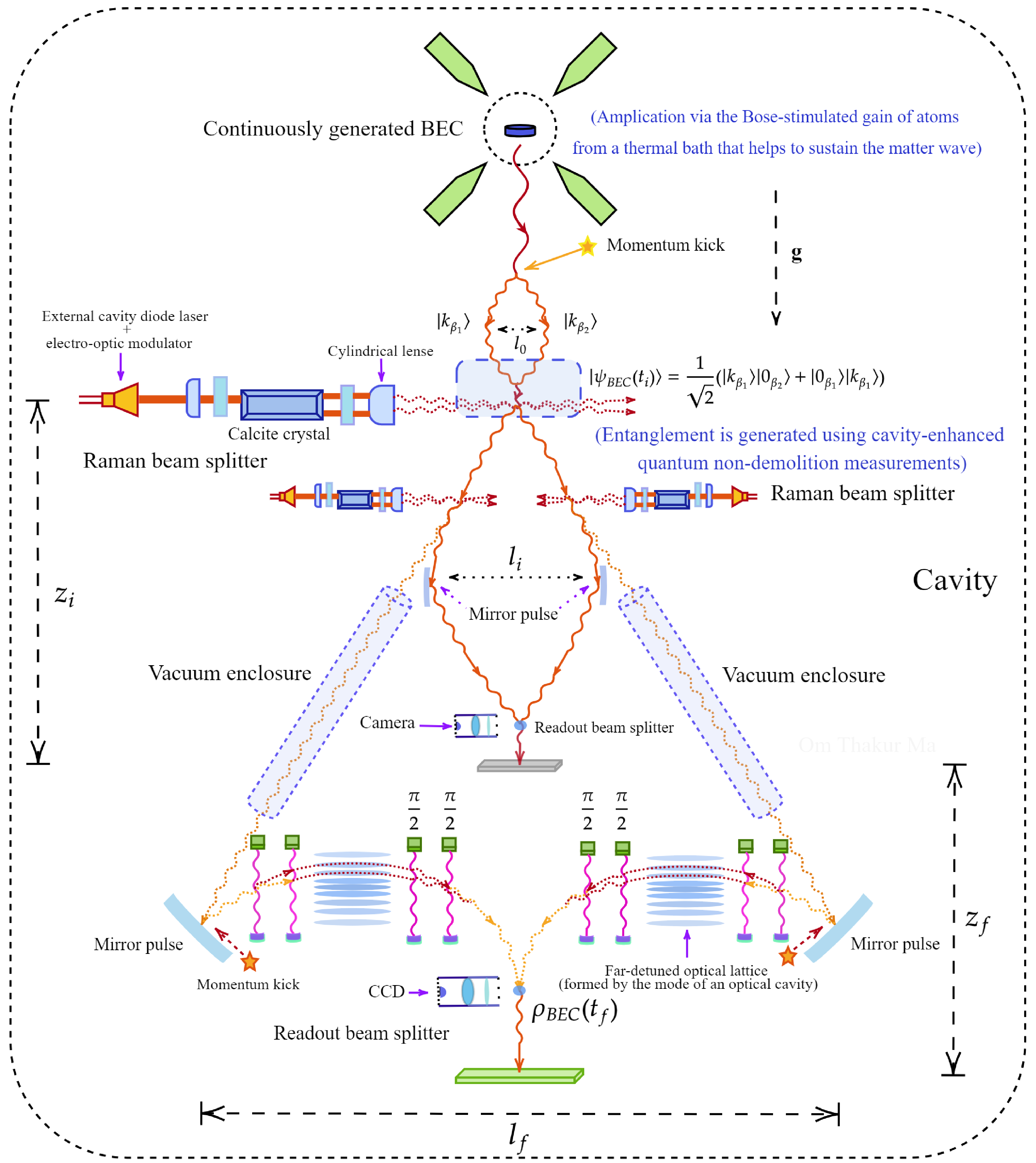}
\caption{A schematic diagram (not to scale) of a BEC-graviton detector placed inside a cavity. The ultra-cold atom laser falls freely under the effect of Earth's gravity and creates an interference pattern twice with a finite time gap between the two interferences. For a graviton signature, one should rely on the coherence loss in the final interference phenomenon compared to the initial one.  \label{Graviton_Detector_final}}
\end{figure}
\end{center}
\end{widetext}

\noindent This helps to create a continuously generated atom laser for a BEC-based atom interferometer. This technique is also widely known as ``output-coupling" \cite{ContinuousBEC}. The next step is to generate a maximally entangled state $|\psi_{\text{BEC}}\rangle$ as in eq.(\ref{2.39}) from the generated atom laser (denoted by a dark red wavy line in Fig.(\ref{Graviton_Detector_final})).

\item \textcolor{blue}{\textit{Creation of a maximally entangled BEC state}}

\noindent Our primary aim is to create a maximally entangled momentum state mimicking the state in eq.(\ref{2.39}). The atom interferometer-based measurements using entangled atomic ensembles allow one to surpass the restrictions employed by the standard quantum limit (SQL). One can use the momentum entanglement techniques in \cite{PRLEntangled2} but our model is mainly based on the entanglement procedure imposed in \cite{NatureEntangled}. The freely falling atom lasers are then divided into two beams by using quantized momentum kicks (generates different momentum in the two atoms laser beams) where the maximum separation between the two is $l_0\sim 10-20 ~\mu$m such that the coherence is not lost between the two beams. The two beams are then further recombined. This splitting and recombination can be implemented by a two-photon Raman transition. Atomic probe light inside the cavity is then used to entangle the atoms via the implementation of quantum non-demolition-based measurement techniques (QND) \cite{nondemolition1,nondemolition2,nondemolition3}. This helps us to create the maximally entangled state as can be seen from Fig.(\ref{Graviton_Detector_final}).
\item \textcolor{blue}{\textit{First phase of interference}}

\noindent In order to create an interference pattern we suggest the use of Raman beam splitters to split the maximally entangled combined atom laser into two coherent beams. In Fig.(\ref{Graviton_Detector_final}), we have drawn a schematic diagram of a Raman beam splitter that uses an external cavity diode laser combined with an electro-optic modulator. The laser is then passed through a calcite crystal and cylindrical lenses which helps to split the maximally entangled atom laser beam. For the diagram of the Raman beam splitter we have followed the schematic representation of the beam splitter in \cite{Beamsplitter}. We propose for implementation of two more Raman beam splitters corresponding to each splitted atom laser beam. This shall result in the creation of two pairs of coherent atom laser beams. The nearest two coherent beams can then be refocused using two mirror pulses \cite{NatureEntangled} separated by a distance $l_i$. The refocused beams then interfere with each other after a vertical distance $z_i$ from the initial splitting point of the maximally entangled atom-laser beam as can be seen from Fig.(\ref{Graviton_Detector_final}). Here, $l_i$ can be of the order of $0.5$ mm. We therefore will suggest keeping $l_i\sim 20-60~\mu$m as in the second face of interference the mirror separation $l_f$ must be greater than $l_i$. Finally, the interference pattern is observed using a readout beam splitter.  We also need to give an estimate of the initial height $z_i$.  As can be found in \cite{MultibeamAtomLaser}, atom lasers generally have a free-fall of $t\approx20$ ms. If the atom-laser sources remain coherent for $20$ ms time then the distance of free fall is about (ignoring external influences like Earth's rotation and gravitational free-fall) $z_{\text{cl}}=\frac{1}{2}gt^2$. If one considers quantum gravitational effects then \cite{AppleParikh,OTMGraviton} $z\sim z_{\text{cl}}\pm\Delta z\sim  z_{\text{cl}}+\sqrt{\cosh(2r_k)}l_p\sim z_{\text{cl}}\pm e^{r_k} l_p$. Even for gravitons with a $r_k=42$ squeezing $\Delta z\sim 10^{-17}$ m which can be easily neglected in our current analysis. For $t\approx 20$ ms, the free-fall distance is $z\approx 2$ mm. We propose to keep $z_i$ in the 40-80$\mu$m regime so that the free-fall time is $t\sim 2-4 $ ms. This time gap will already introduce a $23-24\%$ decoherence in the BEC \textit{supercondensate} for $r_k=42$ with no phonon squeezing. For more clear experimental outcome $z_i$ should be made as small as possible.
\begin{enumerate}
\item \textcolor{blue}{\textit{Graviton interaction before $t_i$}}\\
Another important aspect to keep in mind is that although the graviton starts interacting with the BEC as soon as the background is perturbed by the gravitational fluctuation, we have up to now considered that the graviton state starts interacting soon after the creation of the maximally entangled BEC state. In an experimental scenario, the graviton will start to interact with the system just after the formation of the BEC and even before the formation of the maximally entangled state. It is crucial to remember that the important scenario in consideration is the decoherence from the super condensate which is formed as a result of the graviton interaction with the maximally entangled BEC state. Hence, the interaction of the graviton with the BEC before the formation of the maximally entangled state is inconsequential to our current analysis unless the initial interaction completely decays all the effects induced by the noise of gravitons. Let us try to understand the procedure analytically. Let us say that at $t=t_i-\delta t$, the BEC is prepared and the atom laser falls under the effect of gravity. Then the state of the system reads (provided gravitons have started to interact with it), $|\psi_{\text{S}}(t_i-\delta)\rangle=|k_\beta\rangle\otimes |h,0\rangle.$ For the graviton state $|h,0\rangle$, `zero' denotes the time of interaction of the gravitons with the BEC. If $\delta t$ is the time required for the separation and formation of the maximally entangled momentum state then before the first separation (using Raman beam splitter) the state of the system takes the form $|\psi_{\text{S}}(t_i)\rangle=\frac{1}{\sqrt{N}}(|\textbf{k}_{\beta_1},0_{\beta_2}\rangle\otimes|h,\delta t,\textbf{k}_{\beta_1}\rangle+|\textbf{k}_{\beta_1},0_{\beta_2}\rangle\otimes|h,\delta t,\textbf{k}_{\beta_2}\rangle)$, where $\frac{1}{\sqrt{N}}$ gives the normalization factor. After a time $t_f$ (second phase of interference), the influence functional needs to be carefully calculated where eq.(2.41g), will now be given as $e^{i\Delta (t_f+\delta t)}=\langle h,t_f-t_i+\delta t,\textbf{k}_{\beta_2}|h,t_f-t_i+\delta t,\textbf{k}_{\beta_1}\rangle$. The lower limits of the integrations in eq.(\ref{2.42}) will be replaced by $\delta t$ and the decoherence factor at the final time will then be $\Gamma(t_f+\delta t)$. One can carefully now estimate the changes in the decoherence effect due to this additional $\delta t$ time for minute detailing, instead, we propose to keep $\delta t$ as small as possible to avoid extra decoherence effects in the final interference pattern. One can also consider constructing a scenario where the first part of the experiment (formation of the BEC to the creation of the maximally entangled state) is shielded from incoming gravitational waves but, in reality, it is considered to be an impossible task. Several proposals have been made on how to shield a system from the effects of linearized gravity \cite{GWShielding0,GWShielding1,GWShielding2}. One can also implement stimulated absorption of gravitons through continuous sensing of quantum jumps as been proposed recently in \cite{GravitonAbsorption}. Instead of graviton shielding we tend towards carefully reducing $\delta t$ as it will be more experimentally efficient and easy to control in an experimental setup.
\end{enumerate}
\item \textcolor{blue}{\textit{Second phase of interference}}

\noindent The second phase of interference consists of the other two beams which were not refocused using mirror pulses (denoted by wavy and dotted orange lines in Fig.(\ref{Graviton_Detector_final})). As the decoherence increases with time our primary goal is to increase the total time interval between the initial splitting and second phase of interference $t_f-t_i$. For this part, one can follow three distinct ways. 

\begin{enumerate}
\item \textcolor{blue}{\textit{Use of far detuned optical lattice:}} The best way to increase the time interval $t_f-t_i$ is to make use of far detuned optical lattice laser. At first the split atom lasers (wavy dotted orange lines after the second phase of beam-splitting in Fig.(\ref{Graviton_Detector_final})) are passed through a vacuum enclosure until they are launched into a projectile upwards using mirror pulses and momentum kicks\footnote{The vacuum enclosures are introduced to reduce decoherence due to interaction with other forms of radiation. One can also enclose all such projected trajectories using such vacuum enclosures to reduce any unwanted coherence-loss.}. Here we propose the use of the methodology and technique in \cite{NatureStableAtomLaser}. The atoms are then loaded into the regions with the highest intensities corresponding to the standing wave of an optical lattice laser. This optical lattice laser is formed by the fundamental mode of a Fabry-Perot cavity, oriented in the vertical direction. Two $\pi/2$ pulses are used to redirect and recombine the launched atomic wave packets. In general, interference can be observed at this level itself but one can proceed a step further by interfering with the recombined atomic wave packets from the two set ups which is analyzed using the final read-out beam splitter. The distance between the two mirror pulses (perpendicular distance from the two center points) $l_f$ can be made as high as $500$ $\mu\text{m}$ to  $2$ mm \cite{Beamsplitter}. It is important to note that $l_f$ should not be so large that the coherence between the two atom laser beams is lost. The launched atomic wave packets using the final mirror pulses can sustain their coherence inside the far-detuned optical lattice for as long as 60-70 seconds after which they are again launched from the lattice as has been observed in \cite{NatureStableAtomLaser}. The vertical distance between the first and second interference phenomenon $z_f$ can be set such that $z_f>z_i$ which will further enhance the decoherence effect. For convenience, we set $z_f=2$ mm. We estimate an overall coherence loss of $39\%$ which is equivalent to a $20\%$ of relative coherence loss in the second phase of interference corresponding to the first phase of interference\footnote{We are not taking into effect the decoherence due to atomic ensemble dephasing. For an experimental set up, one needs to carefully take care of the other factors. We are primarily focusing on the decoherence due to bremsstrahlung from the \textit{supercondensate}.}. 
\item \textcolor{blue}{\textit{Free-fall in the second phase:}} The second way of generating the second phase of entanglement is via using free-fall and recombination of the separated atom beams from the second level of beam splitting. The decoherence effect is enhanced only via the increase in the value of $z_f$. This does impose the problem of a larger free-fall distance than the previous case which can in principle introduce an unwanted decoherence effect in the final stage of interference.
\item \textcolor{blue}{\textit{Doppler compensation technique :}} We propose a Doppler compensation technique by making a movable second phase of interference. To do so one can introduce a sliding mechanism containing the bottom mirror pulses (which can be moved horizontally to adjust for the freely falling atom-beams), and the beam splitter. Introducing an optical lattice set up for the Doppler case will be very difficult and as a result, it is easy to progress with the free-fall model. It is possible to give a certain blue shift to the entire moving mechanism such that the coherence loss is compensated by giving a certain upward velocity. If the velocity is such that the coherence loss is compensated, it is possible to calculate the coherence loss due to Bremsstrahlung from the \textit{supercondensate} via calculating the velocity required for the Doppler shift. As it will be very difficult to impart an exact velocity to mitigate the coherence loss we keep it as a future direction for our initial experimental model.
\end{enumerate}
\end{enumerate}
One important thing to note is that one can make use of momentum squeezing techniques \cite{NatureEntangled} to further amplify the viability of such BEC-based atom-interferometers. One can also introduce more interference events by splitting the two split atom beams for multiple times. However, such multiple splitting may result in unavoidable measurement errors in the experiment.

\noindent The only challenge of this experimental set up lies in the fact that without the LISA observatory up in outer space, we cannot be sure whether a primordial gravitational wave has really arrived or not. Significant experimental enhancement can be achieved if the length of the initial BEC is increased to the order of $1$m. If gravitons exist with primordial gravity waves exhibiting such a high squeezing factor then within a decade's time gravitons should be experimentally detected using the model proposed in this analysis. 
\section{Conclusion}\label{S5}
\noindent In this paper, we extend the analysis presented in \cite{Super_Condensate_OTM,
Super_Condensate_OTM_Lett} and analyze the effect of graviton-induced noise on a maximally entangled two-mode state of a weakly coupled Bose-Einstein condensate. We take the total action for the BEC-graviton detector model from \cite{Super_Condensate_OTM,
Super_Condensate_OTM_Lett} and extracting the Lagrangian from it, we obtain the total Hamiltonian for the system. Now, using a density matrix formalism, we make use of the Liouville equation of motion to obtain the solution of the final density matrix of the system in terms of the initial density matrix. Tracing over the gravitational field operators and with enough analytical calculations, we arrive at the most important result in our paper which is eq.(\ref{2.35}). For a BEC system, initially, with a maximally entangled momentum eigenstate, we observe that due to the noise induced by the gravitons a new \textit{supercondensate} is formed where the separable BEC-graviton state becomes mixed in nature. We call it a Bose-Einstein \textit{supercondensate}. We also observe a time-dependent decoherence and an overall entanglement degradation as a result of the gravitational Bremsstrahlung from the \textit{supercondensate} as a result of noise fluctuations due to the gravitons interacting with the BEC. We observe that the exponential decay factor decays significantly with higher values of the graviton squeezing parameter. For a graviton squeezing $r_k=42$, we observe a $10\%$ coherence loss by $2\mu\text{s}$ and the logarithmic negativity of the system decays by $10\%$ in same time interval. The decay of the logarithmic negativity over time confirms a theoretical loss of entanglement which is also amplified by higher graviton squeezing. We also investigated the effect of phonon-squeezing on decoherence and observed a substantial loss of coherence due to a minor increase in the real phonon-squeezing parameter. A similar $10\%$ coherence loss over a time interval of $2\mu$s can be observed for the phonon squeezing $r_\omega=11.0$ whereas the graviton squeezing is only $r_k=20$. This gives a more definite experimental control as the phonon squeezing parameter can be tuned experimentally. This also opens up a way to investigate graviton signatures while the initial gravitational squeezing is not very high.  Finally, we propose an experimental set up that uses a dual Mach-Zehnder interferometer model to detect such coherence loss via interfering maximally entangled coherent atom laser beams continuously generated from a continuous wave Bose-Einstein condensation. The proposed experimental set up is given in Fig.(\ref{Graviton_Detector_final}). This experiment makes use of the idea that with a high enough interaction time of the BEC with the gravitons, the coherence loss in the \textit{supercondensate} state will be more significant with time. As a result, if two interference patterns are compared, which are kept at different heights, the closest one to the ground level will exhibit a higher loss of coherence than the one kept at a higher level. The primary reason behind this observation is that the lower one has a higher time of interaction with the gravitons than the one kept at a higher height which results in a higher coherence loss. One can further increase the time interval between the two interference events by using far-detuned optical lattice up to a minute order following the procedure in \cite{NatureStableAtomLaser}. For such an experiment, one needs to exploit gravitational waves generated during the inflationary time period where there is a high enough possibility for the gravitons being highly squeezed. We hope that with the upcoming LISA observatory, detection of graviton signatures is going to be a matter of just a decade.


\begin{thebibliography}{8}
\bibitem{Singularity_Free_Theories_Gravity}
T. Biswas, E. Gerwick, T. Koivisto, and A. Mazumdar, ``\textit{Towards Singularity- and Ghost-Free Theories of Gravity}", \href{https://link.aps.org/doi/10.1103/PhysRevLett.108.031101}{Phys. Rev. Lett. 108 (2012) 031101}.
\bibitem{Karolyhazy}
F. Karolyhazy, ``\textit{Gravitation and quantum mechanics of macroscopic objects}", \href{https://doi.org/10.1007/BF02717926}{Nuovo Cimento A 42 (1966) 390}.
\bibitem{Penrose}
R. Penrose, ``\textit{On Gravity's role in Quantum State Reduction}", \href{https://doi.org/10.1007/BF02105068}{Gen. Relativ. Gravit. 28 (1996) 581}.
\bibitem{Diosi}
L. Di\'{o}si, ``\textit{Models for universal reduction of macroscopic quantum fluctuations}", \href{https://link.aps.org/doi/10.1103/PhysRevA.40.1165}{Phys. Rev. A 40 (1989) 1165}.
\bibitem{Bassi}
A. Bassi, K. Lochan, S. Satin, T. P. Singh, and H. Ulbricht, ``\textit{Models of wave-function collapse, underlying theories, and experimental tests}", \href{https://link.aps.org/doi/10.1103/RevModPhys.85.471}{Rev. Mod. Phys. 85 (2013) 471}.
\bibitem{Bassi2}
A. Bassi, A. Gro{\ss}ardt, and H. Ulbricht, ``\textit{Gravitational decoherence}", \href{https://iopscience.iop.org/article/10.1088/1361-6382/aa864f}{Class. Quant. Gravit. 34 (2017) 193002}.
\bibitem{Gorelik}
G. E. Gorelik, ``\textit{Matvei Bronstein and quantum gravity: 70th anniversary of the unsolved problem}", \href{https://iopscience.iop.org/article/10.1070/PU2005v048n10ABEH005820}{Phys.-Usp. 48 (2005) 1039}.
\bibitem{Feynman}
R. Feynman, \textit{in Chapel Hill Conference Proceedings, 1957}.
\bibitem{Chapel_Hill}
P. G. Bergman, ``\textit{Summary of the Chapel Hill Conference}", \href{https://link.aps.org/doi/10.1103/RevModPhys.29.352}{Rev. Mod. Phys. 29 (1957) 352}.
\bibitem{Bronstein1}
M. P. Bronstein, ``\textit{Quantentheorie schwacher Gravitationsfelder}", \href{https://edition-open-sources.org/sources/10/22/index.html}{Phys. Z. Sowjetunion 9 (1936) 140}.
\bibitem{Bronstein2}
M. P. Bronstein, ``\textit{Kvantovanie gravitatsionnykh voln}", Zh. Eksp. Teor. Fiz. 6 (1936) 195.
\bibitem{SNGupta1}
S. N. Gupta, ``\textit{Quantization of Einstein's Gravitational Field: Linear Approximation}", \href{https://iopscience.iop.org/article/10.1088/0370-1298/65/3/301}{Proc. Phys. Soc. A 65 (1952) 161}.
\bibitem{SNGupta2}
S. N. Gupta, ``\textit{Quantization of Einstein's Gravitational Field: General Treatment}", \href{https://iopscience.iop.org/article/10.1088/0370-1298/65/8/304}{Proc. Phys. Soc. A 56 (1952) 608}.
\bibitem{BlumRickels}
A. S. Blum and D. Rickels (eds.), ``\textit{Quantum Gravity in the First Half of the Twentieth Century: A Sourcebook}", \href{https://edition-open-sources.org/sources/10/index.html}{Berlin: Max-Planck-Gesellschaft zur Förderung der Wissenschaften (2018)}.
\bibitem{Marletto}
C. Marletto and V. Vedral, ``\textit{Witness gravity’s quantum side in the lab}", \href{https://doi.org/10.1038/547156a}{Nature 547 (2017) 156}.
\bibitem{Bose}
S. Bose, A. Mazumdar, G. W. Morley, H. Ulbricht, M. Toro\v{s}, M. Paternostro, A. A. Geraci, P. F. Barker, M. S. Kim, and G. Milburn, ``\textit{Spin Entanglement Witness for Quantum Gravity}", \href{https://link.aps.org/doi/10.1103/PhysRevLett.119.240401}{Phys. Rev. Lett. 119 (2017) 240401}. 
\bibitem{Marletto2}
C. Marletto and V. Vedral, ``\textit{Gravitationally Induced Entanglement between Two Massive Particles is Sufficient Evidence of Quantum Effects in Gravity}", \href{https://link.aps.org/doi/10.1103/PhysRevLett.119.240401}{Phys. Rev. Lett. 119 (2017) 240402}.
\bibitem{Bose2}
R. J. Marshman, A. Mazumdar, and S. Bose, ``\textit{Locality and entanglement in table-top testing of the quantum nature of linearized gravity}", \href{https://link.aps.org/doi/10.1103/PhysRevA.101.052110}{Phys. Rev. A 101 (2020) 052110}.
\bibitem{Bose3}
S. Bose, A. Mazumdar, M. Schut, and M. Toro\v{s}, ``\textit{Mechanism for the quantum natured gravitons to entangle masses}", \href{https://doi.org/10.1103/PhysRevD.105.106028}{Phys. Rev. D 105 (2022) 106028}.
\bibitem{Bose_Interferometry}
P. Fragolino, M. Schut, M. Toro\v{s}, S. Bose, and A. Mazumdar, ``\textit{Decoherence of a matter-wave interferometer due to dipole-dipole interactions}", \href{https://doi.org/10.1103/PhysRevA.109.033301}{Phys. Rev. A 109 (2024) 033301}.
\bibitem{Haine}
S. A. Haine, ``\textit{Searching for signatures of quantum gravity in quantum gases
}", \href{https://doi.org/10.1088/1367-2630/abd97d}{New J. Phys. 23 (2021) 033020}.
\bibitem{QGravNoise}
M. Parikh, F. Wilczek, and G. Zahariade, ``\textit{The noise of gravitons}", \href{https://www.worldscientific.com/doi/abs/10.1142/S0218271820420018}{Int. J. Mod. Phys. D 29 (2020) 2042001}.
\bibitem{QGravLett}
M. Parikh, F. Wilczek, and G. Zahariade, ``\textit{Quantum Mechanics of Gravitational Waves}", \href{https://link.aps.org/doi/10.1103/PhysRevLett.127.081602}{Phys. Rev. Lett. 127 (2021) 081602}.
\bibitem{QGravD}
M. Parikh, F. Wilczek, and G. Zahariade, ``\textit{Signatures of the quantization of gravity at gravitational wave detectors}", \href{https://link.aps.org/doi/10.1103/PhysRevD.104.046021}{Phys. Rev. D 104 (2021) 046021}.
\bibitem{AppleParikh}
S. Chawla and M. Parikh, ``\textit{Quantum gravity corrections to the fall of an apple}", \href{https://link.aps.org/doi/10.1103/PhysRevD.107.066024}{Phys. Rev. D 107 (2023) 066024}.
\bibitem{OTMGraviton}
S. Sen and S. Gangopadhyay, ``\textit{Minimal length scale correction in the noise of gravitons}, \href{https://doi.org/10.1140/epjc/s10052-023-12230-2}{Eur. Phys. J. C 83 (2023) 1044}.
\bibitem{OTMApple}
S. Sen and S. Gangopadhyay, ``\textit{Uncertainty principle from the noise of gravitons}", \href{https://doi.org/10.1140/epjc/s10052-024-12481-7}{Eur. Phys. J. C 84 (2024) 116}.
\bibitem{KannoSodaTokuda}
S. Kanno, J. Soda, and J. Tokuda, ``\textit{Noise and decoherence induced by gravitons}", \href{https://link.aps.org/doi/10.1103/PhysRevD.103.044017}{Phys. Rev. D 103 (2021) 044017}.
\bibitem{KannoSodaTokuda2}
S. Kanno, J. Soda, and J. Tokuda, ``\textit{Indirect detection of gravitons through quantum entanglement}", \href{https://link.aps.org/doi/10.1103/PhysRevD.104.083516}{Phys. Rev. D 104 (2021) 083516}.
\bibitem{SNBose}
S. N. Bose, ``\textit{Plancks Gesetz und Lichtquantenhypothese}", \href{https://doi.org/10.1007/BF01327326}{Zeitschrift f\"{u}r Physik 26 (1924) 178}.
\bibitem{Einstein1}
A. Einstein, ``\textit{Quantentheorie des einatomigen idealen Gases}", \href{https://doi.org/10.1002/3527608958.ch27}{Sitz. Preu{\ss}. Akad. Wiss. 10 (1924) 261}.
\bibitem{Einstein2}
A. Einstein, ``\textit{Quantentheorie des einatomigen idealen Gases. Zweite Abhandlung.}", \href{https://doi.org/10.1002/3527608958.ch28}{Sitz. Preu{\ss}. Akad. Wiss. 8 (1925) 3}\footnote{Links of the references \cite{Einstein1,Einstein2} refer to chapters 27 and 28 of \cite{Einstein3}.}.
\bibitem{Einstein3}
D. Simon, ``\textit{\textbf{Albert Einstein: Akademie-Vortr\"{a}ge}, Sitzungsberichte der Preu{\ss}ischen Akademie der Wissenschaften 1914–1932}", Wiley-VCH (2006), Berlin.
\bibitem{1Nobel2001}
M. H. Anderson, J. R. Ensher, M .R. Matthews, C. E. Wieman, and E. A. Cornell, ``\textit{Observation of Bose-Einstein Condensation in a Dilute Atomic Vapor}", \href{https://www.science.org/doi/10.1126/science.269.5221.198}{Science 269 (1995) 198}.
\bibitem{2Nobel2001}
K. B. Davis, M.-O. Mewes, M. R. Andrews, N. J. van Druten, D. S. Durfee, D. M. Kurn, and W. Ketterle, ``\textit{Bose-Einstein Condensation in a Gas of Sodium Atoms}", \href{https://link.aps.org/doi/10.1103/PhysRevLett.75.3969}{Phys. Rev. Lett. 75 (1995) 3969}.
\bibitem{PhononBEC}
C. Sab\'{i}n, D. E. Bruschi, M. Ahmadi, and I. Fuentes, ``\textit{Phonon creation by gravitational waves}",  \href{https://iopscience.iop.org/article/10.1088/1367-2630/16/8/085003}{New J. Phys. 16 (2014) 085003}.
\bibitem{PhononBEC2}
R. Sch\"{u}tzhold, ``\textit{Interaction of a Bose-Einstein condensate with a gravitational wave}", \href{https://link.aps.org/doi/10.1103/PhysRevD.98.105019}{Phys. Rev. D 98 (2018) 105019}
\bibitem{PhononBEC3}
M. P. G. Robbins, N. Affshordi, and R. B. Mann, ``\textit{Bose-Einstein condensates as gravitational wave detectors}", \href{https://doi.org/10.1088/1475-7516/2019/07/032}{J. Cosmology Astroparticle Phys. 07 (2019) 032}.
\bibitem{PhononBEC4}
M. P. G. Robbins, N. Affshordi, A. O. Jamison, and R. B. Mann, ``\textit{Detection of gravitational waves using parametric resonance in Bose-Einstein condensates}", \href{https://iopscience.iop.org/article/10.1088/1361-6382/ac7b05}{Class. Quant. Gravit. 39 (2022) 175009}.
\bibitem{ThesisMatthew}
M. P. G. Robbins, ``\textit{Quantum Information across Spacetime: From Gravitational Waves to Spinning Black Holes}", Ph.D. Thesis, University of Waterloo (2021).
\bibitem{Super_Condensate_OTM_Lett}
S. Sen and S. Gangopadhyay, ``\textit{Bose-Einstein condensate as a quantum gravity probe}; ``\textbf{\textit{Erste Abhandlung}}", \href{https://doi.org/10.48550/arXiv.2404.06060}{arXiv:2404.06060 [hep-th]}.
\bibitem{Super_Condensate_OTM}
S. Sen and S. Gangopadhyay, ``\textit{Probing the quantum nature of gravity using a Bose-Einstein condensate}, \href{https://link.aps.org/doi/10.1103/PhysRevD.110.026014}{Phys. Rev. D 110 (2024) 026014}.
\bibitem{DymnikovaKhlopov}
I. Dymnikova and M. Khlopov, ``\textit{Decay of cosmological constant as Bose condensate evaporation}", \href{https://doi.org/10.1142/S0217732300002966}{Mod. Phys. Lett. A 15 (2000) 2305}.
\bibitem{DymnikovaKhlopov2}
I. Dymnikova and M. Khlopov, ``\textit{Decay of cosmological constant in self-consistent inflation}", \href{https://link.springer.com/article/10.1007/s100520100625}{Eur. Phys. J. C 20 (2001) 139}.
\bibitem{SpaceBased}
B. Wang, B. Li, Q. Xiao, G. Mo, and Y.-F. Cai, ``\textit{Space-based optical lattice clocks as gravitational wave detectors in search for new physics}", \href{https://doi.org/10.48550/arXiv.2410.04340}{arXiv:2410.04340 [gr-qc]}. 
\bibitem{bremsstrahlung}
H.-P. Breuer and F. Petruccione, ``Destruction of quantum coherence through emission of bremsstrahlung", \href{https://link.aps.org/doi/10.1103/PhysRevA.63.032102}{Phys. Rev. A 63 (2001) 032102}.
\bibitem{LinWolfe}
J. l. Lin and J. P. Wolfe, ``\textit{Bose-Einstein Condensation of Paraexcitons in Stressed $\text{Cu}_2\text{O}$}", \href{https://link.aps.org/doi/10.1103/PhysRevLett.71.1222}{Phys. Rev. Lett. 71 (1993) 1222}.
\bibitem{Chelkowski}
S. Chelkowski, H. Vahlbruch, B. Hage, A. Franzen, N. Lastzka, K. Danzmann, and R. Schnabel, ``\textit{Experimental characterization of frequency-dependent squeezed light}", \href{https://link.aps.org/doi/10.1103/PhysRevA.71.013806}{Phys. Rev. A 71 (2005) 013806}.
\bibitem{Johnsson}
M. T. Johnsson, G. R. Dennis, and J. J. Hope, ``Squeezing in Bose-Einstein condensates with a large number of atoms", \href{https://dx.doi.org/10.1088/1367-2630/15/12/123024}{New. J. Phys. 15 (2013) 123024}.
\bibitem{GuLiWuYang}
W. Gu, G. Li, S. Wu, and Y. Yang, ``\textit{Generation of non-classical states of mirror motion in the single-photon strong-coupling regime}", \href{https://doi.org/10.1364/OE.22.018254}{Opt. Express 22 (2014) 18254}. 
\bibitem{FuentesMann}
I. Fuentes-Schuller and R. B. Mann, ``Alice Falls into a Black Hole: Entanglement in Noninertial Frames", \href{https://link.aps.org/doi/10.1103/PhysRevLett.95.120404}{Phys. Rev. Lett. 95 (2005) 120404}.
\bibitem{3Nobel2001}
W. Ketterle, ``\textit{Nobel lecture: When atoms behave as waves: Bose-Einstein condensation and the atom laser}", \href{https://link.aps.org/doi/10.1103/RevModPhys.74.1131}{Rev. Mod. Phys. 74 (2002) 1131}.
\bibitem{AtomInterferometry}
S. Dimopoulos, P. W. Graham, J. M. Hogan, M. A. Kasevich, and S. Rajendran,  ``\textit{Gravitational wave detection with atom interferometry}", \href{https://doi.org/10.1016/j.physletb.2009.06.011}{Phys. Lett. B 678 (2009) 37}.
\bibitem{AtomInterferometry2}
J. M. Hogan \textit{et. al.}, ``\textit{An atomic gravitational wave interferometric sensor in low earth orbit (AGIS-LEO)}", \href{https://link.springer.com/article/10.1007/s10714-011-1182-x}{Gen. Relativ. Gravit. 43 (2011) 1953}.
\bibitem{DualbeamAtomLaser}
N. Lundblad, R. J. Thompson, D. C. Aveline, and L. Maleki, ``\textit{Spinor dynamics-driven formation of a dual-beam atom laser}", \href{https://doi.org/10.1364/OE.14.010164}{Opt. Exp. 14 (2006) 10164}.
\bibitem{MultibeamAtomLaser}
J. Dugu\'{e}, G. Dennis, M. Jeppesen, M. T. Johnsson, C. Figl, N. P. Robbins, and J. D. Close, ``\textit{Multibeam atom laser: Coherent atom beam splitting from a single far-detuned laser}, \href{https://link.aps.org/doi/10.1103/PhysRevA.77.031603}{Phys. Rev. A 77 (2008) 031603(R)}.
\bibitem{NatureEntangled}
G. P. Grev, C. Luo, B. Wu, and J. K. Thompson, ``\textit{Entanglement-enhanced matter-wave interferometry in a high-fineness cavity}", \href{https://doi.org/10.1038/s41586-022-05197-9}{Nature 610 (2022) 472}.
\bibitem{NatureEntangled2}
J. Esteve, C. Gross, A. Weller, S. Giovanazzi, and M. K. Oberthaler, ``Squeezing and entanglement in a Bose-Einstein condensate", \href{https://doi.org/10.1038/nature07332}{Nature 455 (2008) 1216}.
\bibitem{NatureEntangled3}
C. Gross, T. Zibold, E. Nicklas, J. Esr\'{e}ve, and M. K. Oberthaler, ``\textit{Nonlinear atom interferometer surpasses classical precision limit}", \href{https://doi.org/10.1038/nature08919}{Nature 464 (2010) 1165}.
\bibitem{NatureEntangled4}
R. B\"{u}cker, J. Grond, S. Manz, T. Berrada, C. Koller, U. Hohenester, T. Schumm, A. Perrin, and J. Schmiedmayer, ``\textit{Twin-atom beams}", \href{https://doi.org/10.1038/nphys1992}{Nat. Phys. 7 (2011) 608}.
\bibitem{NatureEntangled5}
C. D. Hamley, C. S. Gerving, T. M. Hoang, E. M. Bookjans, and M. S. Chapman, ``\textit{Spin-nematic squeezed vacuum in a quantum gas}", \href{https://doi.org/10.1038/nphys2245}{Nat. Phys. 8 (2012) 305}.
\bibitem{ScienceEntangled}
X.-Y. Luo, Y.-Q. Zou, L.-N. Wu, Q. Liu, M.-F. Han, M. K. Tey, and L. You, ``\textit{Deterministic entanglement generation from driving through quantum phase transitions}", \href{https://doi.org/10.1126/science.aag1106}{Science 355 (2017) 620}.
\bibitem{ScienceEntangled2}
M. Fadel, T. Zibold, B. D\'{e}camps, and P. Treutlein, ``\textit{Spatial entanglement patterns and Einstein-Podolsky-Rosen steering in Bose-Einstein condensates}", \href{https://doi.org/10.1126/science.aao1850}{Science 360 (2018) 409}.
\bibitem{ScienceEntangled3}
K. Lange, J. Peise B. L\"{u}cke, I. Krise, G. Vitagliano, I. Apellaniz, M. Kleinmann, G. T\'{o}th, and C. Klempt, ``\textit{Entanglement between two spatially separated atomic modes}", \href{https://doi.org/10.1126/science.aao2035}{Science 360 (2018) 416}.
\bibitem{NatureEntangled6}
D. Leibfried, E. Knill, S. Seidelin, J. Britton, R. B. Blakestad, J. Chiaverini, D. B. Hume, W. M. Itano, J. D. Jost, C. Langer, R. Ozeri, R. Reichle, and D. J. Wineland, ``\textit{Creation of a six-atom `Schr\"{o}dinger cat' state}", \href{https://doi.org/10.1038/nature04251}{Nature 438 (2005) 639}.
\bibitem{PRLEntangled}
T. Monz, P. Schindler, J. T. Barreiro, M. Chwalla, D. Nigg, W. A. Coish, M. harlander, W. H\"{a}nsel, M. Hennrich, and R. Blatt, ``\textit{14-Qubit Entanglement: Creation and Coherence}", \href{https://link.aps.org/doi/10.1103/PhysRevLett.106.130506}{Phys. Rev. Lett. 106 (2011) 130506}.
\bibitem{PRLEntangled2}
F. Anders, A. Idel, P. Feldmann, D. Bondarenko, S. Loriani, K. Lange, J. Peise, M. Gersemann, B. Meyer-Hoppe, S. Abend, N. Gaaloul, C. Schubert, D. Schlippert, L. Santos, E. Rasel, and C. Klempt, ``\textit{Momentum Entanglement for Atom Interferometry}", \href{https://link.aps.org/doi/10.1103/PhysRevLett.127.140402}{Phys. Rev. Lett. 127 (2021) 140402}.
\bibitem{PRLGravWave}
P. W. Graham, J. M. Hogan, M. A. Kasevich, and S. Rajendran, ``\textit{New Method for Gravitational Wave Detection with Atomic Sensors}", \href{http://dx.doi.org/10.1103/PhysRevLett.110.171102}{Phys. Rev. Lett. 110 (2013) 171102}.
\bibitem{PRAGravWave}
M. A. Norcia, J. R. K. Cline, and J. K. Thompson, ``\textit{Role of atoms in atomic gravitational-wave detectors}", \href{https://doi.org/10.1103/PhysRevA.96.042118}{Phys. Rev. A 96 (2017) 042118}.
\bibitem{BoseGraviton}
P. Fragolino, M. Schut, M. Toro\v{s}, S. Bose, and A. Mazumdar, ``\textit{Decoherence of a matter-wave interferometer due to dipole-dipole interactions}", \href{https://link.aps.org/doi/10.1103/PhysRevA.109.033301}{Phys. Rev. A 109 (2024) 033301}.
\bibitem{ContinuousBEC}
C.-C. Chen, R. Gonz\'{a}lez Escudero, J. Min\'{a}\v{r}, B. Pasquiou, S. Bennetts, and F. Schreck, ``\textit{Continuous Bose-Einstein condensation}", \href{https://doi.org/10.1038/s41586-022-04731-z}{Nature 606 (2022) 683}.
\bibitem{nondemolition1}
O. Hosten, N. J. Engelsen, R. Krishnakumar, and M. A. Kasevich, ``\textit{Measurement noise 100 times lower than the quantum-projection limit using entangled atoms}", \href{https://doi.org/10.1038/nature16176}{Nature 529 (2016) 505}.
\bibitem{nondemolition2}
K. C. Cox, G. P. Greve, J. M. Weiner, and J. K. Thompson, ``\textit{Deterministic Squeezed States with Collective Measurements and Feedback}", \href{https://link.aps.org/doi/10.1103/PhysRevLett.116.093602}{Phys. Rev. Lett. 116 (2016) 093602}.
\bibitem{nondemolition3}
S. Colombo, E. Pedrozo-Pe\~{n}afiel, A. F. Adiyatullin, Z. Li, E. Mendez, C. Shu, and V. Vuleti\'{c}, ``\textit{Time-reversal-based quantum metrology with many-body entangled states}", \href{https://doi.org/10.1038/s41567-022-01653-5}{Nat. Phys. 18 (2022) 925}.
\bibitem{Beamsplitter}
D. D\"{o}ring, J. E. Debs, N. P. Robins, C. Figl, P. A. Altin, and J. D. Close, ``\textit{Ramsey interferometry with an atom laser}", \href{https://doi.org/10.1364/OE.17.020661}{Opt. Express 18 (2009) 20661}.
\bibitem{GWShielding0}
W. Ning, ``\textit{Gravitational Shielding Effect in Gauge Theory of Gravity}", \href{https://iopscience.iop.org/article/10.1088/0253-6102/41/4/567}{Commun. Theor. Phys. 41 (2004) 567}.
\bibitem{GWShielding1}
L. A. Savrov, ``\textit{Gravitational Shielding and  the Equivalence Principle}", \href{https://link.springer.com/article/10.1134/S0202289312040081}{Gravitation and Cosmology 18 (2012) 270}.
\bibitem{GWShielding2}
R. Beig and P. T. Chru\'{s}ciel, ``\textit{Shielding linearized gravity}", \href{ https://doi.org/10.1103/PhysRevD.95.064063}{Phys. Rev. D 95 (2017) 064063}.
\bibitem{GravitonAbsorption}
G. Tobar, S. K. Manikandan, T. Beitel, and I. Pikovski, ``\textit{Detecting single gravitons with quantum sensing}", \href{https://www.nature.com/articles/s41467-024-51420-8}{Nat. Commun. 15 (2024) 7229}.
\bibitem{NatureStableAtomLaser}
C. D. Panda, M. Tao, J. Egelhoff, M. Ceja, V. Xu, and H. M\"{u}ller, ``\textit{Coherence limits in lattice atom interferometry at the one-minute scale}", \href{https://doi.org/10.1038/s41567-024-02518-9}{Nat. Phys. 20 (2024) 1234}.
\end{thebibliography}
\end{document}